\documentclass[letterpaper]{article}
\usepackage[preprint]{aaai2027}  
\usepackage[hyphens]{url}  
\usepackage{graphicx} 
\usepackage{natbib}  
\usepackage{caption} 
\usepackage{algorithm}

\usepackage{newfloat}
\usepackage{listings}
\DeclareCaptionStyle{ruled}{labelfont=normalfont,labelsep=colon,strut=off} 
\usepackage{booktabs}

\usepackage{algorithmicx}
\usepackage{tabularx}
\usepackage{booktabs}
\usepackage{amssymb}
\usepackage{placeins}
\usepackage{subcaption}
\usepackage[table]{xcolor}
\definecolor{bestgreen}{HTML}{87E5AA}
\usepackage{xcolor}

\usepackage{enumitem}
\usepackage{adjustbox}
\usepackage{array}
\usepackage{multirow}
\usepackage{adjustbox}
\usepackage{makecell}
\usepackage{amsmath,amssymb,amsfonts}
\usepackage{textcomp}
\usepackage{xcolor}
\usepackage{lipsum}
    
\usepackage{booktabs} 
\usepackage{pifont}
\usepackage{mathtools}

\usepackage{lipsum}
\usepackage{dblfloatfix}
\usepackage{algpseudocode}
\usepackage{bm}
\usepackage{multirow}
\usepackage{verbatim}
\usepackage{listings}
\usepackage{soul}

\definecolor{cadmiumgreen}{rgb}{0.0, 0.42, 0.24}

\newcommand{\drop}[1]{\textcolor{red}{#1}}
\renewcommand{\drop}[1]{}

\lstdefinelanguage{Verilog}{
  morekeywords={module, endmodule, input, output, reg, wire, always, begin, end, if, else, for, while, case, default},
  sensitive=false,
  morecomment=[l]{//},
  morecomment=[s]{/*}{*/},
  morestring=[b]",
}

\definecolor{shadecolor}{rgb}{0.9,0.9,0.9}
\usepackage{latexsym}

\usepackage[utf8]{inputenc} 
\usepackage[T1]{fontenc}    
\usepackage{url}            
\usepackage{booktabs}       
\usepackage{amsfonts}       
\usepackage{nicefrac}       
\usepackage{microtype}      
\usepackage{xcolor}         

\usepackage{tcolorbox}
\tcbuselibrary{skins,breakable} 
\usepackage{listings}
\usepackage{xurl}

\usepackage[utf8]{inputenc}
\usepackage[T1]{fontenc}
\usepackage{listings}
\usepackage{multirow}
\usepackage{pifont}
\usepackage{colortbl}

\newif\ifshowjw
\showjwtrue

\newif\ifshowjv
\showjvtrue

\newif\ifshowkevin
\showkevintrue      

\newcommand{\ourtool}{\texttt{EDATracer}}
\newcommand{\bc}{\cellcolor{bestgreen}}

\title{\ourtool: An Agentic Framework for Large-Scale EDA Artifact Analysis}

\author{
    Phat Tieu,
    Sayanti Jana,
    Matthew DeLorenzo,
    Jiawen Wu,\\
    Narendran Srinivasan,
    Srinivas Shakkottai,
    Jiang Hu,
    Jeyavijayan Rajendran
}
\affiliations{
    Texas A\&M University, Department of Electrical and Computer Engineering\\
    College Station, Texas 77840, USA\\
    \{kevin.tieu, sayantijana008, matthewdelorenzo, realjw, narrow.end.run, sshakkot, jianghu, jv.rajendran\}@tamu.edu
}

\begin{document}

\maketitle

\begin{abstract}
Modern chip design relies on electronic design automation (EDA) tools that generate large, heterogeneous artifacts, including source files, scripts, logs, netlists, and reports. Analyzing these artifacts is critical for debugging, optimization, and design-flow understanding, but remains difficult because relevant evidence is often distributed across many artifact types and design stages. Although LLM agents show promise for EDA assistance, existing approaches lack public benchmarks for large-scale cross-artifact analysis and often struggle to ground reasoning in tool-generated evidence.

We present \ourtool{}, an agentic framework for evidence-grounded EDA artifact analysis. \ourtool{} organizes design artifacts into a knowledge graph paired with a semantic vector index, enabling LLM agents to retrieve evidence across source files, logs, netlists, and reports. We curate an 18.9 GB dataset of 2{,}787 synthesizable open-source chip designs and introduce a 90-question benchmark spanning factual, statistical, and reasoning tasks. Across evaluated agents, \ourtool{} achieves the best pass@1 accuracy, outperforming Cursor and Claude Code by 6.4 and 7.2 percentage points on average, while using \(2.0\text{--}3.2\times\) fewer tokens.
\end{abstract}

\section{Introduction}

Semiconductor chips have become strategic infrastructure for Artificial Intelligence (AI), data centers, and the broader digital economy~\cite{Commerce2024CHIPSTwoYears, SIA2026GlobalSemiconductorSales}. At the core of the chip industry lies electronic design automation (EDA) tools~\cite{siemens_new_era_ai_eda_2026}. Chip design flows use EDA workflows and tools to automate the design, simulation, verification, synthesis, and physical implementation of modern hardware~\cite{Siemens2026EDADefinition}.

Effective \textit{EDA artifact analysis} is essential within the design, optimization, and verification stages~\cite{Groeneveld2002PhysicalDesignChallenges}. Modern chip design flows generate large, heterogeneous collections of artifacts, including up to thousands of log files and millions of lines of output, making analysis increasingly costly and complex~\cite{huang21mlforeda,kanagal2025llm,arneberg2022chewlogs}. This creates significant engineering and economic burdens: artifact analysis accounts for 47\% of chip verification effort, while debugging can delay design schedules in an industry where advanced chip costs have risen from approximately \$298M at 7 nm in 2018 to \$700M at 2 nm in 2026~\cite{foster20222022,palma2022growing,pwc_semiconductor_beyond_2026}. Consequently, EDA artifact analysis remains a practical bottleneck for modern chip design~\cite{chen2026reportnsfworkshopai,arneberg2022chewlogs}.

Large language model (LLM)-based \textit{agentic frameworks} have shown success in assisting with chip design flow~\cite{linkedin_cadence_ai_agents_2026}. 
For example, agentic frameworks have been used for verification assistance, where an LLM helps engineers check whether a chip design behaves as intended~\cite{liu2024rtlcoder, pu2024ragedadocqa}; automated debugging, where an LLM helps identify the root cause of design or tool failures~\cite{li2025edadebugger}; and EDA workflow automation, where an LLM-based agent coordinates multi-step design flow tasks such as invoking tools, interpreting intermediate results, and selecting next actions~\cite{lu2026autoeda}. 
These frameworks typically combine retrieval, external analysis tools, and multi-step reasoning to solve complex technical tasks~\cite{chen2026maeda,rag2020neurips,yao2023react,shinn2023reflexion}.

\begin{figure}[t]
    \centering
    \includegraphics[width=1\linewidth]{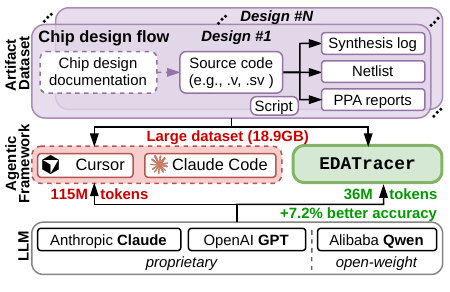}
    \caption{Overview of \ourtool{}, an agentic framework for EDA artifact analysis.}
    \label{fig:teaser}
\end{figure}

However, EDA artifact analysis poses unique challenges for existing LLM and agentic frameworks due to 3 key factors:
\textbf{C1: Lack of public datasets and benchmarks.} Previous EDA datasets target chip design or logic synthesis~\cite{liu2024rtlcoder,yu2026selfevolvedabc}. In contrast, EDA artifacts contain sensitive design details, tool configurations, and project-specific debugging information. Therefore, there is no publicly available benchmark that systematically supports cross-artifact EDA analysis. Furthermore, no widely used open benchmark currently exists to evaluate agents' capabilities to accurately analyze EDA artifacts in large-scale design repositories~\cite{kanagal2025llm,chen2026reportnsfworkshopai}.
\textbf{C2: Complex EDA artifacts.} EDA analysis requires reasoning over many designs, where each design may contain numerous heterogeneous and lengthy artifacts. For example, a single design can produce multiple types of artifacts, including source code, documentation, logs, netlists, and reports. Each artifact can exceed $100k$ lines in large designs. This makes naive full-context prompting impractical due to context and token limits~\cite{qi2023loggpt}. 
\textbf{C3: Grounding failures in cross-artifact reasoning.} EDA artifact analysis often requires evidence from multiple files and design stages; e.g., the final failure message appears in the log file, but the actual source of the failure may be hidden in a related source file or in an earlier report. Without a reliable reference of the EDA-tool context to ground LLM reasoning to real evidence, LLMs may hallucinate or rely on shallow keyword matches~\cite{pu2024ragedadocqa}. 

We propose \textbf{\ourtool{}} (Figure~\ref{fig:teaser}), an agentic framework for evidence-grounded analysis of large chip design repositories and their EDA artifacts. To address dataset limitations in scale and EDA-related artifacts (\textbf{C1}), we develop an automated dataset curation methodology to extract, filter, and synthesize chip repositories, yielding 2{,}787 large-scale designs along with their synthesis logs, netlists, and performance reports. To address the size and heterogeneity of EDA artifacts (\textbf{C2}), we organize the dataset into a \textit{knowledge graph} representation that captures design hierarchy, artifact provenance, and dependency relationships. To support grounded cross-artifact reasoning (\textbf{C3}), \ourtool{} uses a retrieval agent to identify relevant evidence and a reasoning agent to synthesize answers from the retrieved context. Across a 90-question benchmark covering factual, statistical, and reasoning tasks, \ourtool{} outperforms existing commercial agent frameworks while using fewer tokens.

In summary, the contributions of this paper are:
\begin{itemize}[leftmargin=*, itemsep=0pt, topsep=0pt]

\item We curate an 18.9 GB dataset with 2,787 open-source, large-scale designs and associated EDA artifacts capturing the design hierarchy and cross-artifact dependencies.

\item We introduce \ourtool{}, an agent framework that organizes EDA artifacts into a knowledge graph and uses structured retrieval with LLM reasoning for grounded question answering.

\item We develop a 90-question benchmark for EDA artifact analysis that covers factual, statistical, and reasoning tasks at varying levels of difficulty.
 
\item We provide a comprehensive evaluation against state-of-the-art agent frameworks, \ourtool{} achieves the best overall accuracy, outperforming Cursor and Claude Code pass@1 by an average of 6.4 and 7.2 percentage points, respectively, while also using up to $3.2\times$ fewer total tokens.

\end{itemize}

\section{Background}
\label{sec:background}
\noindent \textbf{Electronic design automation (EDA)} tools are a suite of software and hardware tools that assist engineers across various stages of the chip design flow, automating complex tasks required to design, verify, and manufacture chips~\cite{Groeneveld2002PhysicalDesignChallenges, huang21mlforeda}. 
For instance, EDA tools can assist at the initial design specification stage, in which developers specify the intended functionality of a chip with code\footnote{In this paper, code and source code for chip design refer to register-transfer level (RTL) hardware description language (HDL) code, such as Verilog and SystemVerilog.}. 
Logic \textit{synthesis} tools then transform code into low-level representations (gate-level netlists), ensuring the design can be implemented using an existing technology library and providing estimates of power, performance, and chip area (PPA)~\cite{synopsys_design_compiler, cadence_genus}. 
EDA tools also assist at different stages of chip design (e.g., simulation, manufacturing, and fabrication)~\cite{Siemens2026EDADefinition}. 

\noindent \textbf{EDA artifacts.} EDA tools produce artifacts essential for developers to analyze in support of their design objectives~\cite{kanagal2025llm}. 
These artifacts include the (i) \textit{synthesis script}, which invokes the synthesis tool and specifies the design to be synthesized, the technology library to use, the constraints to apply, and the synthesis outputs to generate.
(ii) \textit{Synthesis logs}, which track the detailed compilation procedure from code to netlists, identifying execution status (e.g., design parsing, elaboration, and technology mapping), errors, and warnings (e.g., syntax issues, missing files/modules, or bad practices)~\cite{arneberg2022chewlogs}. 
(iii) \textit{Netlist} files are also generated, detailing the exact logic gates, registers, and interconnections required to manufacture the associated chip design~\cite{synopsys_design_compiler, cadence_genus,wang2024nextgenlogic}. 
Additionally, (iv) \textit{PPA reports} are generated post-synthesis, which calculate the estimated power consumed, the performance (timing), and the physical area of the chip~\cite{huang21mlforeda}. These quantities are central optimization targets in recent learning-based EDA systems~\cite{xue2024macroregulator, wang2025symrtlo}.

\noindent \textbf{EDA artifact analysis.} 
Existing LLM-based analysis methods remain largely artifact-local. 
Log analysis targets software logs for anomaly detection or  diagnosis~\cite{qi2023loggpt,liu24logprompt}. 
Chip-oriented LLM studies focus on code generation~\cite{bush2025freefairhardware} or report-level prediction~\cite{wang2024chippowereda}. 
Another work evaluates multi-document QA benchmarks over evidence distributed across documents~\cite{tang2024multihoprag}. 
However, these tasks lack EDA-specific artifact semantics and design-flow dependencies.
To the best of our knowledge, prior work lacks systematic, grounded analysis over linked chip design artifacts. \ourtool{} addresses this gap by organizing EDA artifacts into a structured cross-artifact representation and enabling grounded multi-step reasoning across the design flow.

\begin{figure}[t]
    \centering
    \includegraphics[width=1\linewidth]{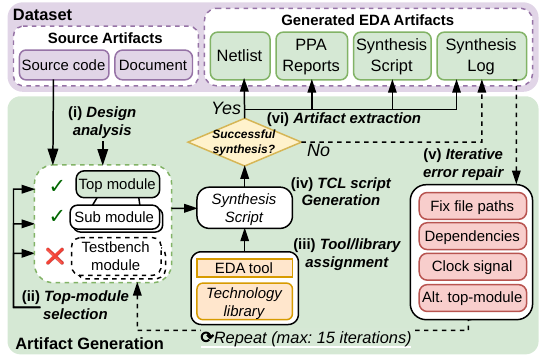}
    \caption{EDA artifact dataset construction flow.}
    \label{fig:dataset_flowchart}
\end{figure}

\section{Creating the Dataset}
\label{sec:dataset_creation}

A representative dataset for agent-based EDA analysis must pair diverse, synthesizable chip designs with the EDA artifacts generated during the design flow. 
This pairing is necessary because realistic EDA tasks require the capacity to reason across both designs and tool-generated artifacts, where relevant evidence is often distributed across multiple files. 
Existing EDA datasets are limited for this setting, as many focus on code generation~\cite{chen2026maeda} or single-log analysis~\cite{bush2025freefairhardware} rather than cross-artifact reasoning. 

To this end, we construct our EDA artifact dataset with two goals: (i) collecting diverse, full-chip open-source designs (\textit{source artifacts}), and (ii) generating the pairing comprehensive flow artifacts (\textit{generated artifacts}) using common EDA tools and libraries~\cite{synopsys_design_compiler,cadence_genus}. 
 Curating our initial dataset has two stages: the \textit{collection stage} and the \textit{generation stage}.

\subsection{Collection Stage}
\label{sec:design_collection}

The collection stage constructs a corpus $\mathcal{D}$ of open-source chip designs (\textit{source artifacts}) from public GitHub repositories.
These repositories are often inconsistently organized, and not synthesis-ready. 
Our goal is to retain synthesizable chip designs spanning diverse categories, including processors, memories, and controllers. We formulate design collection as a staged filtering process that progressively refines an initial candidate set into a curated corpus.
We represent each candidate repository as $r=(\mathcal{S}_r,\mathcal{M}_r)$, where $\mathcal{S}_r$ is the repository source-file set and $\mathcal{M}_r$ contains repository metadata, such as the repository name, description, and directory structure. 
If a repository is accepted into the final corpus, it becomes a design $d\in\mathcal{D}$ with source-file set $\mathcal{S}_d$.

Let $\mathcal{K}$ denote a set of chip-design-related search keywords.
The collection process proceeds in four phases:
(i) \textbf{Discovery}, which queries public repositories using $\mathcal{K}$ and forms an initial candidate set
$\mathcal{R}_0 = \bigcup_{k \in \mathcal{K}} \textsc{Search}(k)$
after deduplication at the URL-level.
(ii) \textbf{Heuristic filtering}, which applies rule-based predicates $h(r)$ over repository files and metadata to remove candidates unlikely to contain usable RTL, yielding
$\mathcal{R}_1 = \{r \in \mathcal{R}_0 : h(r)=1\}$.
(iii) \textbf{LLM-based evaluation}, which applies a fixed-prompt evaluator $\ell(r)$ to assess whether each remaining repository corresponds to a meaningful chip design with potentially synthesizable structure, yielding
$\mathcal{R}_2 = \{r \in \mathcal{R}_1 : \ell(r)=1\}$.
(iv) \textbf{Synthesis feasibility}, which applies a synthesis check $v(r)$ and retains only repositories that can be processed by the downstream artifact-generation pipeline:
$\mathcal{D} = \{d_r : r \in \mathcal{R}_2,\ v(r)=1\}$,
where $d_r$ denotes the accepted design from repository $r$.

The accepted designs form a curated corpus for EDA artifact generation. 
Further implementation details can be found in the Appendix.

\subsection{Generation Stage}
\label{sec:artifact_generation}


\begin{algorithm}[t]
\footnotesize
\caption{Iterative Repair Loop}
\label{alg:iterative_repair_intext}
\begingroup
\newcommand{\AlgI}{\hspace{1.4em}}
\newcommand{\AlgII}{\hspace{2.8em}}
\begin{algorithmic}[1]
\Require Top-module candidates $\mathcal{T}$, HDL files $\mathcal{V}$, clock candidates $\mathcal{C}$, max attempts $A_{\max}$
\Ensure Synthesized artifacts or terminal failure

\State $\mathcal{T}_{\mathrm{tried}} \gets \emptyset$
\State $a \gets 0$
\State $t \gets \mathrm{top}(\mathcal{T})$
\State $c \gets \mathrm{best}(\mathcal{C})$
\State $S \gets \textsc{MakeTCL}(t,\mathcal{V},c)$

\State \textbf{while} $a < A_{\max}$ \textbf{do}
    \State \AlgI $a \gets a+1$
    \State \AlgI $\mathcal{T}_{\mathrm{tried}} \gets \mathcal{T}_{\mathrm{tried}} \cup \{t\}$
    \State \AlgI $(L,e,R) \gets \textsc{RunSynthesis}(S)$

    \State \AlgI \textbf{if} \textsc{Complete}$(L)$ \textbf{and} \textsc{NonZeroArea}$(R)$ \textbf{then}
        \State \AlgII Store log, netlist, and PPA reports
        \State \AlgII \Return \textsc{Success}
    \State \AlgI \textbf{end if}

    \State \AlgI $\gamma \gets \textsc{ClassifyAttempt}(L,e,R)$
    \State \AlgI $\Delta \gets \textsc{SelectUpdate}(\gamma,\mathcal{T},\mathcal{T}_{\mathrm{tried}},\mathcal{C})$

    \State \AlgI \textbf{if} $\Delta = \textsc{NoRepair}$ \textbf{then}
        \State \AlgII \Return \textsc{Failure}$(\gamma)$
    \State \AlgI \textbf{end if}

    \State \AlgI $S \gets \textsc{ApplyUpdate}(S,\Delta)$
    \State \AlgI $(t,c) \gets \textsc{UpdateState}(t,c,\Delta)$
    \State \AlgI \textsc{LogAttempt}$(t,\gamma,\Delta)$
\State \textbf{end while}

\State \Return \textsc{Failure}(\textsc{AttemptLimit})
\end{algorithmic}
\endgroup
\end{algorithm}

The objective of the artifact generation stage is to synthesize each accepted design $d\in\mathcal{D}$ and produce a generated EDA artifact tuple $\mathcal{A}_d = (L_d, N_d, P_d)$,
where $L_d$ denotes the synthesis log, $N_d$ the gate-level netlist, and $P_d$ the set of PPA reports. 
Here, $\mathcal{S}_d$ denotes the original source-file set associated with design $d$, while $\mathcal{A}_d$ denotes the artifacts generated by our synthesis flow. 
Because public repositories typically do not include complete synthesis artifacts or synthesis-ready configurations, we formulate artifact generation as a deterministic, iterative synthesis procedure that maps each design either to a valid artifact tuple $\mathcal{A}_d$ or to a terminal failure state. 
As shown in Figure~\ref{fig:dataset_flowchart}, artifact generation proceeds through the following six stages for each design $d$.

\noindent \textbf{(i) Tool and library assignment} assigns a synthesis configuration $\theta_d=(\tau_d,\lambda_d)$, where $\tau_d$ is the synthesis tool and $\lambda_d$ is the technology library.
\footnote{We use Synopsys Design Compiler~\cite{synopsys_design_compiler} as the synthesis tool and three open-source technology libraries: NanGate45~\cite{nangate45_lib}, SkyWater130~\cite{skywater130_lib}, and ASAP7~\cite{asap7_lib}. Library selection will be recorded in the generated artifacts.}

\noindent \textbf{(ii) Design analysis} applies a static analyzer to the repository file set $\mathcal{S}_d$ to identify candidate HDL files $\mathcal{V}_d$, remove likely testbench or verification files, reconstruct the module-instantiation graph $H_d$, rank candidate top modules $\mathcal{T}_d$, and detect candidate clock signals $\mathcal{C}_d$ when possible. Details regarding HDL file selection and top-module ranking are provided in the Appendix.

\noindent \textbf{(iii) Top-module selection} initializes synthesis with the highest-ranked candidate $t_d \in \mathcal{T}_d$, which is necessary because many repositories do not explicitly specify a synthesizable top module.

\noindent \textbf{(iv) TCL script generation} constructs a tool-specific synthesis script $S_d$ from $(\mathcal{V}_d,t_d,\mathcal{C}_d,\theta_d)$. If a clock signal is detected, the script includes an explicit timing constraint; otherwise, it introduces a virtual clock to enable timing analysis.

\noindent \textbf{(v) Iterative error repair} repeatedly invokes synthesis and updates the script configuration based on the observed failure category, as detailed in Algorithm~\ref{alg:iterative_repair_intext}. 
Each failed attempt is classified using the synthesis log and report, and at most one deterministic update is applied per iteration: switching to the next top-module candidate, extending the search path, or selecting an alternative clock. 
The loop terminates when synthesis succeeds, no valid update remains, or the max attempt count is reached.

\noindent \textbf{(vi) Artifact extraction} stores $\mathcal{A}_d=(L_d,N_d,P_d)$ for each successful synthesis run. 
The final artifact dataset therefore consists of successfully synthesized designs, their original source-file sets $\mathcal{S}_d$, and their generated artifacts $\mathcal{A}_d$. 
For later knowledge-graph construction, we define the complete ingestion set for each design as
$\mathcal{I}_d=\mathcal{S}_d\cup\{L_d,N_d\}\cup P_d$.

\begin{figure}[t]
    \centering
    \includegraphics[width=1.0\linewidth]{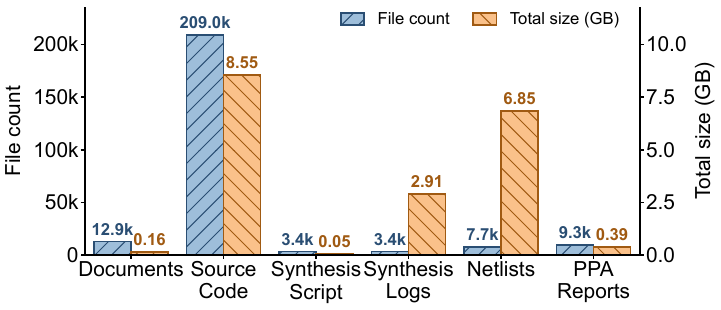}
    \caption{Distribution of file types in the dataset.}
    \label{fig:dataset_artifact_test}
\end{figure}

\subsection{Dataset Results}

We obtain the following results. 
The collection stage first identified 23,115 candidate designs, of which 16,514 passed heuristic filtering, 8,812 passed LLM-based evaluation, and 2,787 synthesizable full-chip designs were accepted into the final EDA artifact dataset.
The distribution of chip design sources is included in the Appendix.
Notably, the largest category included RISC-V architectures (23.6\%).

After the generation stage, artifacts (synthesis logs, netlists, and PPA reports) for each of the top-level designs were generated, as shown in Figure~\ref{fig:dataset_artifact_test}. In total, we obtained a dataset of 18.91 GB, of which 53\% is attributable to EDA-generated artifacts, supplementing the curated chip designs.

\section{\ourtool{} Agentic Framework}
\label{sec:framework}
\begin{figure*}[t]
    \centering
    \includegraphics[width=1\textwidth]{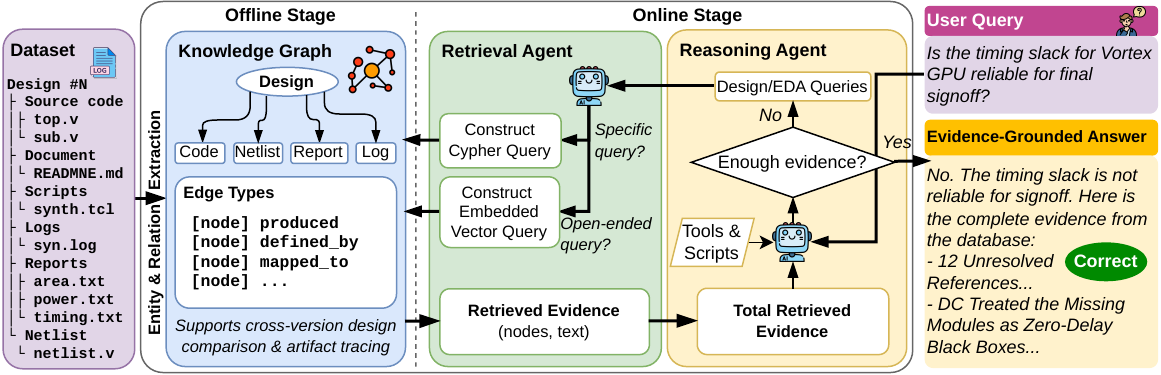}
    \caption{The full \ourtool{} agentic framework.}
    \label{fig:framework_figure}
\end{figure*}

The goal of \ourtool{} is to answer a user question $q$ by identifying and reasoning over relevant evidence distributed across the design ingestion sets $\{\mathcal{I}_d\}_{d\in\mathcal{D}}$. 
This setting creates two challenges: (i) artifacts are too large and heterogeneous for direct full-context prompting (\textbf{C2}), and (ii) many questions require evidence that spans multiple files and artifact types (\textbf{C3}). 
As shown in Figure~\ref{fig:framework_figure}, \ourtool{} addresses these challenges through offline and online stages. 
In the \textit{offline stage}, \ourtool{} converts each design's ingestion set into a structured knowledge graph and a graph-associated semantic vector index. 
In the \textit{online stage}, a retrieval agent and reasoning agent interact over these representations to produce grounded answers with supporting evidence.

\subsection{Offline Stage}
\label{sec:offline_stage}

For each design $d$, \ourtool{} maps the ingestion set $\mathcal{I}_d$ into two complementary representations: an attributed, typed knowledge graph $G_d=(V_d,E_d)$ and a semantic vector index $\mathcal{Z}_d$. 
For the knowledge graph, $V_d$ and $E_d$ denote the set of graph nodes and directed typed edges, respectively. 
Each node $v\in V_d$ represents an EDA-relevant entity, such as a design, artifact file, text chunk, RTL module, netlist component, or structural element. 
Node-specific information is stored as metadata $\phi(v)$, including artifact type, file path, line range, textual content, extracted metrics, warning or error labels, and synthesis status. 
Each edge is represented as $(u,\rho,v)$, where $u,v\in V_d$ and $\rho$ denotes the relation type. 
Edges encode artifact hierarchy, provenance, and structural relations.

Also, \ourtool{} builds a semantic vector index over text-bearing graph nodes:
$\mathcal{Z}_d=\{(v,z_v):v\in V_{d,\mathrm{emb}}\subseteq V_d\}$,
where $V_{d,\mathrm{emb}}$ contains nodes selected for semantic retrieval, such as text-chunk nodes, and $z_v$ is the embedding associated with node $v$. 
The embedding vectors are stored in the vector index rather than as ordinary node metadata in $\phi(v)$, while the corresponding graph node preserves source information such as the design identifier, artifact path, and line span. 
The full offline representation is
$G=(V,E), \quad V=\bigcup_{d\in\mathcal{D}}V_d,\quad E=\bigcup_{d\in\mathcal{D}}E_d,
\qquad
\mathcal{Z}=\bigcup_{d\in\mathcal{D}}\mathcal{Z}_d$.

Graph ingestion consists of three core operations:
(i) \textbf{Artifact parsing} partitions input files $\mathcal{I}_d$ into textual files $\mathcal{I}_d^{\mathrm{text}}$ (e.g., logs, reports) and structural files $\mathcal{I}_d^{\mathrm{struct}}$ (e.g., RTL, netlists), using domain-specific parsers to extract node metadata $\phi(v)$.
(ii) \textbf{Provenance linking} adds typed edges to preserve structural hierarchy and lineage across generated EDA artifacts.
(iii) \textbf{Semantic indexing} splits long text files into overlapping chunks, instantiating chunk nodes in $G_d$ and indexing their embeddings in $\mathcal{Z}_d$. 
This enables semantic retrieval to locate relevant nodes via the vector index and map them back to their graph context $G$. 
The complete procedure is detailed in the supplementary material.

\subsection{Online Stage}

At online inference time, \ourtool{} answers a user question $q$ using the knowledge graph $G=(V,E)$ and graph-associated vector index $\mathcal{Z}$. 
The online stage maps
$(q,G,\mathcal{Z}) \rightarrow (\hat{y},\mathbb{E}_q)$,
where $\hat{y}$ is the generated answer and
$\mathbb{E}_q=(V_q,E_q,\mathcal{B}_q)$
is the accumulated evidence. 
Here, $V_q\subseteq V$ and $E_q\subseteq E$ denote retrieved graph nodes and edges, while $\mathcal{B}_q\subseteq\mathcal{Z}$ denotes retrieved vector-index hits. 
As each vector-index entry has the form $(v,z_v)$, each semantic hit is associated with a graph node whose metadata $\phi(v)$ provides the relevant text span, file path, lines, extracted EDA facts, and context.

As shown in Figure~\ref{fig:framework_figure}, \ourtool{} addresses \textbf{C2} and \textbf{C3} through an iterative interaction between a \textit{reasoning agent} and a \textit{retrieval agent}. 
The reasoning agent interprets the user question, determines required information, and generates information requests. 
The retrieval agent answers these requests using semantic search over $\mathcal{Z}$ or structured Cypher queries~\cite{neo4j_cypher_query_language} over $G$. 
The reasoning agent then integrates the returned evidence, and either produces an answer or requests additional evidence from the retrieval agent.

\noindent\textbf{Retrieval agent.}
At iteration $i$, let $\mathcal{R}^{(i)}=\{r_1^{(i)},\ldots,r_m^{(i)}\}$ denote the information requests generated by the reasoning agent. 
For each request $r_j^{(i)}$, the retrieval agent can use two retrieval modes.

In \textit{semantic retrieval}, the agent embeds the request and searches the vector index with cosine similarity:
$\mathcal{B}_{\mathrm{sem}}(r_j^{(i)})
=
\operatorname{TopK}_{(v,z_v)\in\mathcal{Z}}
\mathrm{cos\_sim}\bigl(\textsc{Embed}(r_j^{(i)}),z_v\bigr).$
The returned entries identify graph-associated text nodes, such as chunks from logs, reports, documentation, or scripts, which can be resolved through $G$ to recover their design and artifact context.

In \textit{structured retrieval}, the agent generates a Cypher query $c_j^{(i)}$ and executes it over the knowledge graph:
$(\Delta V_{\mathrm{cyp}},\Delta E_{\mathrm{cyp}})
=
\textsc{ExecCypher}(c_j^{(i)},G).$
This retrieves graph evidence directly, such as design nodes, artifact nodes, extracted metrics, structural relations, and provenance links. 
The semantic and structured results are merged into an evidence update
$\Delta\mathbb{E}^{(i)}=(\Delta V^{(i)},\Delta E^{(i)},\Delta\mathcal{B}^{(i)})$,
which is added to the accumulated evidence:
$\mathbb{E}_q^{(i+1)}
=
\mathbb{E}_q^{(i)}\cup\Delta\mathbb{E}^{(i)}.$

\noindent\textbf{Reasoning agent.}
The reasoning agent coordinates the online analysis. 
Given the question $q$ and accumulated evidence $\mathbb{E}_q^{(i)}=(V_q^{(i)},E_q^{(i)},\mathcal{B}_q^{(i)})$, it either produces an answer or information requests:
\[
\textsc{Reason}(q,\mathbb{E}_q^{(i)})
\rightarrow
\begin{cases}
\hat{y}, & \text{if sufficient evidence},\\
\mathcal{R}^{(i)}, & \text{otherwise}.
\end{cases}
\]
As each vector hit is associated with a graph node, define
$V_{\mathcal{B}}^{(i)}=\{v:(v,z_v)\in\mathcal{B}_q^{(i)}\}.$
The reasoning agent then operates over a prompt representation
$\mathcal{C}_q^{(i)}
=
\textsc{Prompt}\bigl(
V_q^{(i)}\cup V_{\mathcal{B}}^{(i)},\,
E_q^{(i)},\,
\phi
\bigr),$
which serializes selected nodes, their metadata, and their typed relations into the LLM context. 
The reasoning agent then assesses whether this context is sufficient to answer the question. 
If the evidence is insufficient, it generates additional information requests; otherwise, it returns the final grounded answer $\hat{y}$.

\section{Implementation}
\label{sec:experiments}

\begin{table*}[t]
    \centering
    \small
    \setlength{\tabcolsep}{1mm}
    \begin{tabular}{ll*{12}{c}}
        \toprule
        \multirow{2}{*}{\textbf{Model}} 
        & \multirow{2}{*}{\textbf{Tool}} 
        & \multicolumn{3}{c}{\textbf{Factual}} 
        & \multicolumn{3}{c}{\textbf{Statistical}} 
        & \multicolumn{3}{c}{\textbf{Reasoning}} 
        & \multicolumn{3}{c}{\textbf{Overall}} \\
        \cmidrule(lr){3-5} 
        \cmidrule(lr){6-8} 
        \cmidrule(lr){9-11} 
        \cmidrule(lr){12-14}
        & 
        & \textbf{Avg.} & \textbf{pass@1} & \textbf{pass@5}
        & \textbf{Avg.} & \textbf{pass@1} & \textbf{pass@5}
        & \textbf{Avg.} & \textbf{pass@1} & \textbf{pass@5}
        & \textbf{Avg.} & \textbf{pass@1} & \textbf{pass@5} \\
        \midrule
    
        \multirow{3}{*}{Claude Opus 4.7}
        & Cursor      & 9.40 & 86.7 & 90.0 & 8.33 & 66.7 & 83.3 & 7.93 & 74.0 & 96.7 & 8.56 & 75.8 & 90.0 \\
        & Claude Code & 9.33 & 90.3 & 96.7 & 8.40 & 69.3 & 76.7 & 8.00 & 85.3 & \bc 100 & 8.58 & 81.7 & 91.1 \\
        & \textbf{\ourtool{}}  & \bc 9.87 & \bc 97.3 & \bc 100 & \bc 9.57 & \bc 97.3 & \bc 100 & \bc 9.20 & \bc 88.7 & 96.7 & \bc 9.54 & \bc 94.4 & \bc 98.9 \\
        \midrule
    
        \multirow{3}{*}{Claude Sonnet 4.6}
        & Cursor      & \bc 9.53 & 92.7 & \bc 96.7 & 9.00 & 82.0 & 96.7 & \bc 8.40 & 82.7 & \bc 96.7 & 8.98 & 85.8 & \bc 96.7 \\
        & Claude Code & \bc 9.53 & \bc 95.3 & \bc 96.7 & 9.03 & \bc 93.3 & \bc 100 & 7.80 & 76.7 & 90.0 & 8.79 & 88.4 & 95.6 \\
        & \textbf{\ourtool{}}  & 9.40 & 90.7 & 93.3 & \bc 9.70 & 92.0 & \bc 100 & 8.30 & \bc 85.3 & 93.3 & \bc 9.13 & \bc 89.3 & 95.6 \\
        \midrule
    
        \multirow{3}{*}{Claude Haiku 4.5}
        & Cursor      & \bc 9.70 & \bc 96.7 & \bc 100 & 7.67 & 69.3 & 96.7 & 7.10 & \bc 60.7 & 76.7 & 8.16 & 75.6 & \bc 91.1 \\
        & Claude Code & 8.70 & 91.3 & 96.7 & 7.40 & 76.7 & 86.7 & 5.80 & 48.0 & 73.3 & 7.30 & 72.0 & 85.6 \\
        & \textbf{\ourtool{}}  & 8.83 & 85.3 & 93.3 & \bc 9.53 & \bc 94.7 & \bc 100 & \bc 7.67 & 60.0 & \bc 80.0 & \bc 8.68 & \bc 80.0 & \bc 91.1 \\
        \midrule
    
        \multirow{2}{*}{GPT 5.4}
        & Cursor      & 9.27 & 83.3 & 86.7 & 7.93 & \bc 74.0 & \bc 86.7 & 7.07 & 71.3 & 86.7 & 8.09 & 76.2 & 86.7 \\
        & \textbf{\ourtool{}}  & \bc 9.53 & \bc 94.0 & \bc 100 & \bc 8.53 & 73.3 & \bc 86.7 & \bc 8.87 & \bc 80.7 & \bc 96.7 & \bc 8.98 & \bc 82.7 & \bc 94.4 \\
        \midrule
    
        \multirow{2}{*}{GPT 5.4 mini}
        & Cursor      & \bc 8.40 & \bc 80.0 & \bc 93.3 & 6.93 & 57.3 & 70.0 & \bc 7.10 & 55.3 & 83.3 & \bc 7.48 & \bc 64.2 & 82.2 \\
        & \textbf{\ourtool{}}  & 8.07 & 70.0 & \bc 93.3 & \bc 7.30 & \bc 59.3 & \bc 70.0 & 6.83 & \bc 60.0 & \bc 90.0 & 7.40 & 63.1 & \bc 84.4 \\
        \midrule
      
        \multirow{1}{*}{Qwen 3.5-9B}
        & \textbf{\ourtool{}}  & 8.03 & 69.3 & 83.3 & 8.00 & 70.0 & 90.0  & 6.07 & 36.0 &  73.3 & 7.37 & 58.9 & 82.2 \\
    
        \bottomrule
    \end{tabular}%
    \caption{Accuracy comparison between \ourtool{} and other commercial AI coding assistants. Avg. scores are reported on a 0--10 scale; pass@1 and pass@5 are percentages.}
    \label{tab:comparison_accuracy}
\end{table*}

\noindent \textbf{Hardware stack.} Experiments use a server with an NVIDIA H200 GPU (143~GB), AMD EPYC 9554 64-core processors, and 1.1~TB RAM.

\noindent \textbf{Software stack.} We implement \texttt{Memgraph} 3.8.0~\cite{memgraph,memgraph_vector} for knowledge graph storage. \texttt{Neo4j} 5.11 to perform vector search~\cite{neo4j_vector_search}. \texttt{Strands Agents} 1.23.0~\cite{strands} for agent orchestration. 
We compare \ourtool{} with two commercial agentic frameworks: Cursor~\cite{cursor} and Claude Code~\cite{claude_code}. 
We test on different LLMs, from proprietary models of varying sizes, including \texttt{Claude Opus 4.7}, \texttt{Claude Sonnet 4.6}, and \texttt{GPT-5.4}, to smaller models, \texttt{Claude Haiku 4.5}, \texttt{GPT-5.4-mini}, and an open-weight model, \texttt{Qwen 3.5-9B}. 

\noindent \textbf{Evaluation benchmark.}
We manually construct a benchmark with 90 golden question-answer (QA) pairs from the dataset. 
It contains three categories: factual, statistical, and reasoning, with 30 questions each. 
Each category has three difficulty levels: easy, medium, and hard, with 10 questions per level. Details are provided in the Appendix.

\noindent \textbf{Evaluation metrics.}
We use three metrics to evaluate the correctness and efficiency of EDA artifact analysis. 
(i) \textbf{Average score}. Each generated answer receives a score in $[0,10]$ based on the golden answer, and scores are averaged within each category. Human evaluators score each answer against the golden answer on a $[0,10]$ scale, following the detailed rubric in the Appendix.
(ii) \textbf{Pass@k}~\cite{chenEvaluatingLargeLanguage2021} is computed for $k=1$ and $k=5$, averaged over all questions in a category. A question passes if at least one score is 8 or higher. 
(iii) \textbf{Token usage} is computed as the total number of input and output tokens consumed by each agentic framework across five repeated runs of the 90-question benchmark.

\section{Results}
\label{sec:results}
We focus on three research questions (RQs):
\noindent \textbf{RQ1: Accuracy of analysis.} Does \ourtool{} outperform commercial agentic frameworks on EDA artifact analysis tasks? 
\noindent \textbf{RQ2: Efficiency of token usage.} How cost-efficient is \ourtool{} when processing complex, large-scale EDA artifacts?
\noindent \textbf{RQ3: Local deployment feasibility.} Can \ourtool{} enable smaller, locally deployed open-weight models to perform EDA artifact analysis effectively?

\subsection{RQ1: Accuracy of Analysis}
The accuracy results from the evaluations of \ourtool{} relative to alternative agents are provided in Table~\ref{tab:comparison_accuracy}. 
Regarding \textbf{RQ1}, \ourtool{} improves the overall average score over Cursor in four of the five shared LLM settings, with the exception of \texttt{GPT-5.4-mini}. 
Averaged across these settings, \ourtool{} improves over Cursor by +0.49 score points, from 8.25 to 8.75. 
Compared with Claude Code, \ourtool{} improves the overall average score across all three shared Claude settings, with an average gain of +0.89 score points, from 8.22 to 9.12. 
\ourtool{} achieves the best pass@1 accuracy, outperforming Cursor and Claude Code by 6.4 and 7.2 percentage points on average.

Overall, \ourtool{} performs most effective with stronger models. Paired with \texttt{Opus 4.7}, it achieves the highest overall performance, with 94.4\% pass@1 and 98.9\% pass@5.
Compared with other agentic frameworks, one key insight is that the largest gains are in statistical questions. \ourtool{} achieves the highest average score across the five proprietary LLM settings (8.92/10). 
In contrast, simple factual questions show smaller gains, and weaker models benefit less.
With \texttt{Haiku 4.5} model, \ourtool{} improves statistical-question accuracy by 1.86 points over the next-best agent. 
This reflects \ourtool{}'s ability to retrieve targeted evidence from multiple artifacts and perform multi-step aggregation while preserving correct units and prefixes.

\subsection{RQ2: Efficiency of Token Usage}

\begin{figure}[t]
        \centering
        \includegraphics[width=1\linewidth]{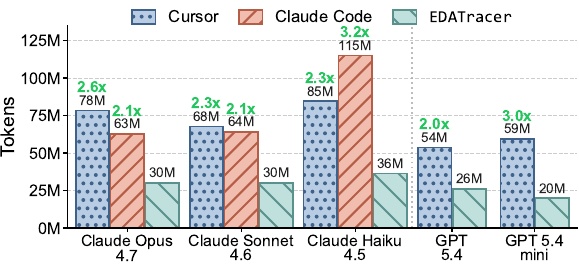}
        \caption{Token usage comparison between \ourtool{} and other commercial agentic frameworks.}
        \label{fig:comparison_token}
    \end{figure}

Figure~\ref{fig:comparison_token} compares the token usage
\footnote{Cursor does not expose detailed token usage through its command-line interface; we obtain its token counts from the aggregate usage statistics reported on the Cursor website.}
of \ourtool{} with Cursor and Claude Code. \ourtool{} consistently uses fewer tokens across all evaluated LLMs, using only 20M--36M tokens compared with 54M--85M for Cursor and 63M--115M for Claude Code. Compared with Cursor, \ourtool{} reduces token usage by 2.0$\times$--3.0$\times$ across five models. Compared with Claude Code, \ourtool{} reduces token usage by 2.1$\times$--3.2$\times$ across the three Claude models. These results show that \ourtool{} reduces token cost for both general-purpose coding agents and model-specific agentic frameworks.

A key insight is that the token reduction is consistent across both Claude and GPT backbones, suggesting that the benefit is derived from \ourtool{}'s retrieval structure rather than from a specific LLM. These results additionally demonstrate that the structured knowledge graph assists in reducing unnecessary context expansion during retrieval and reasoning. Instead of repeatedly passing large artifact contexts to the model, \ourtool{} retrieves targeted graph nodes and typed relationships, allowing the agent to assemble focused evidence for each question. 
Thus, \ourtool{} achieves competitive accuracy while using substantially fewer tokens, demonstrating that graph-based retrieval can improve the cost efficiency of LLM-based EDA artifact analysis.

\subsection{RQ3: Local Deployment Feasibility}
To assess if \ourtool{} can support accurate EDA artifact analysis with a locally deployed open-weight model, we analyze the results from \texttt{Qwen 3.5-9B} (see Table~\ref{tab:comparison_accuracy}).
Although Qwen does not match the strongest proprietary models overall, its performance approaches that of \texttt{GPT-5.4-mini} with \ourtool{} and remains relatively strong on factual and statistical questions.
These results indicate that \ourtool{}’s knowledge-graph representation can provide sufficiently grounded evidence for a local model to answer many artifact-level and aggregation-oriented questions without relying on proprietary inference APIs.

The main limitation appears in reasoning-heavy questions.
Qwen is less successful within a single attempt, although multiple attempts continue to improve overall performance. This pattern suggests that the local model is able to retrieve useful evidence, but has more difficulty than larger commercial models integrating it into a final answer.

\section{Related Work}
\label{sec: related_work}

\noindent \textbf{Chip design datasets}, recently, mainly focus on chip design or logic synthesis tasks, such as RTL generation~\cite{Thakur2023VeriGen}, repair~\cite{liu2024rtlcoder}, and synthesis optimization~\cite{yu2026selfevolvedabc}. 
These datasets are useful for evaluating code-level design tasks, but do not include the full set of artifacts produced during an EDA flow. 
\ourtool{} addresses this gap by building a large-scale dataset of EDA artifacts with linked design-flow artifacts.

\noindent \textbf{EDA agentic frameworks}, recently,
support Verilog generation and repair~\cite{Thakur2023VeriGen,liu2024rtlcoder,blocklove2025autochip}, unit-test generation~\cite{nandal2026laudellmassistedunittest}, EDA question answering~\cite{pu2024ragedadocqa}, root-cause analysis~\cite{qiu25llmbasedrootcauseanalysis}, and workflow automation~\cite{lu2026autoeda}. 
These systems show the promise of LLMs and agents for EDA but target localized code tasks, document QA, or tool execution. 
Meanwhile, graph and KG-based methods support relational reasoning~\cite{Deng2025,anokhin25agrigraph,DeLong_2025}, but EDA graph methods often focus on circuit-level structures~\cite{ma2020understandgrapheda,lopera21gnnforeda,gautam25aipowerneuromorphic}, while graph-based log analysis is more common in software and security~\cite{payne24loganomaly,cotti2025ontologx}. 
\ourtool{} bridges these directions by combining agentic reasoning with graph-grounded retrieval across heterogeneous EDA artifacts.

\section{Conclusion}

We present \ourtool{}, an agentic framework for large-scale EDA artifact analysis. 
We curate an 18.9 GB dataset with 2,787 open-source synthesizable designs and 249K EDA files, including source code, scripts, logs, netlists, reports, and documentation. 
\ourtool{} organizes these artifacts into a domain-specific knowledge graph and combines structured retrieval with LLM reasoning for grounded EDA queries.
We build a benchmark of 90 EDA-focused questions across factual, statistical, and reasoning tasks. 
\ourtool{} achieves the best overall pass@1 accuracy, outperforming Cursor and Claude Code by an average of 6.4\% and 7.2\%, while using between 2.0-3.2$\times$ fewer tokens.

\noindent \textbf{Future work.}
\ourtool{} currently focuses on synthesis-stage artifacts. 
Our future work will cover design-flow stages, such as floorplanning, placement, routing, timing closure, and sign-off, which introduce artifacts including placement reports, congestion maps, timing reports, and sign-off summaries. 
This would enable the knowledge graph to capture more complete design-flow dependencies and support queries that trace timing violations to their sources, including source code, constraints, placement, or routing congestion. 
Another direction is multimodal EDA analysis: future versions can index and reason about visual artifacts such as layouts, floor plans, waveforms, and schematics using vision-language models and multimodal embeddings.

\section{Limitations}\label{sec:limitations}
The current framework contains the following limitations. First, the accuracy of \ourtool{} depends heavily on the quality of knowledge graph construction. If artifacts are poorly parsed, incorrectly chunked, or missing important graph relationships, downstream retrieval and reasoning quality can degrade. In particular, errors in entity and relation extraction can cause the agent to retrieve incomplete or misleading context. This limitation is especially important for EDA workflows, where a warning message, timing value, or report summary is only useful if it is correctly linked to the corresponding design, tool run, script, module, or generated output. Incorrect or missing graph edges can therefore directly affect the reliability of multi-hop retrieval and reasoning.

Second, \ourtool{} introduces an offline preprocessing cost because artifacts must be collected, parsed, embedded, and inserted into the knowledge graph before retrieval can occur. While this cost enables more structured and traceable analysis, it may be less suitable for very small or rapidly changing projects, where a simple file search suffices. In addition, the current schema and benchmark focus primarily on digital EDA artifacts, including source code, synthesis scripts, tool logs, generated reports, netlists, and PPA data. As a result, the framework may require additional engineering to support broader end-to-end chip design workflows, including analog design, physical design, verification, sign-off, and proprietary industrial flows with vendor-specific artifact formats.

\section{Acknowledgment}
The authors acknowledge the support from the Purdue Center for Secure Microelectronics Ecosystem — CSME\#210205.

\bibliography{references}

\clearpage

\section{Technical Appendices and Supplementary Material}
\subsection{EDA Artifact Dataset and Artifact Details}
\label{app:dataset-artifacts}

\begin{figure}[t]
    \centering
    \includegraphics[width=1.0\linewidth]{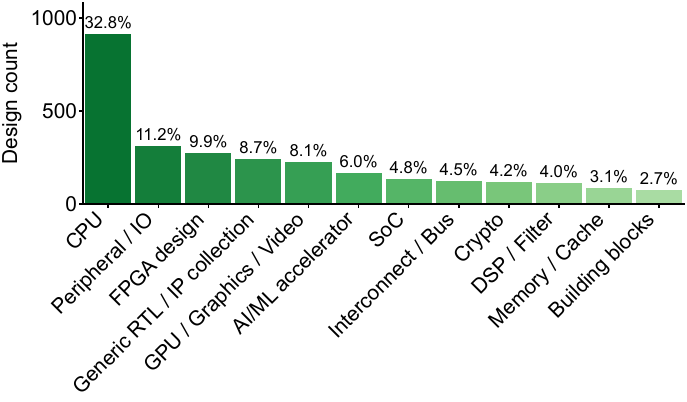}
    \caption{Distribution of chip designs in the dataset.}
    \label{fig:dataset_design}
\end{figure}

This section provides additional context for the EDA artifact dataset introduced in the dataset creation section. It complements the dataset creation process by characterizing what the dataset contains: diversity of collected designs in Figure~\ref{fig:dataset_design} and artifact types associated with each design in Table~\ref{tab:eda_artifacts}.

Figure~\ref{fig:dataset_design} represents the distribution of design categories assigned during the design collection stage described in the design collection. The largest category consists of CPU/processor designs, reflecting the prevalence of open-source processor cores, such as RISC-V designs, in public RTL repositories. The remaining designs span across peripherals, RTL/IP blocks, GPU/graphics blocks, AI/ML accelerators, SoC integrations, interconnect \& bus fabrics, cryptographic blocks, DSP/filter blocks, memory/cache designs, and miscellaneous building blocks. This distribution enables the dataset for EDA artifact analysis across varied RTL structures, design hierarchies, and synthesis outcomes.

Table~\ref{tab:eda_artifacts} summarizes the six artifact categories used in the dataset: (i) documentation, (ii) source code, (iii) synthesis scripts, (iv) synthesis logs, (v) netlists, and (vi) PPA reports. These artifacts capture complementary views of the same design. Documentation describes the design intent; source code defines the RTL implementation; synthesis scripts record tool invocations and setup; synthesis logs capture tool execution; netlists represent the synthesized gate-level implementation; and PPA reports summarize post-synthesis power, performance, and area estimates. These artifacts provide the evidence needed for grounded analysis of EDA artifacts.

The EDA artifact dataset is useful for \ourtool{} because its artifacts are both heterogeneous and connected. They differ in format, including natural-language documentation, HDL source code, TCL scripts, textual logs, gate-level netlists, and tabular reports. Yet, they remain linked by the synthesis process: scripts configure tool execution, logs record what happened during the run, netlists capture the generated implementation, and PPA reports summarize the resulting design quality. This motivates the structured artifact representation described in the framework section, in which \ourtool{} can reason about related artifacts rather than treating each file as an isolated document.

\begin{table*}[t]
    \centering
    \small
    \setlength{\tabcolsep}{1mm}
    \begin{tabular}{
    >{\centering\arraybackslash}m{0.14\linewidth}
    >{\centering\arraybackslash}m{0.18\linewidth}
    >{\centering\arraybackslash}m{0.23\linewidth}
    >{\centering\arraybackslash}m{0.35\linewidth}}
    \toprule
    \textbf{Artifact Type} & \textbf{Example Files} & \textbf{Role in Chip Design Flow} & \textbf{Use in EDA Artifact Analysis} \\
    \midrule
    Documentation 
    & \makecell[c]{\texttt{README.md}\\\texttt{*.md}\\\texttt{*.txt}}
    & Describes design intent, supported features, configuration assumptions, integration notes, and expected behavior. 
    & Provides human-readable context for interpreting whether the synthesized design matches the intended functionality or configuration. \\
    \midrule
    Source code 
    & \makecell[c]{\texttt{*.v}\\\texttt{*.sv}\\\texttt{*.vh}\\\texttt{*.svh}}
    & Defines the register-transfer-level (RTL) chip design, including modules, datapaths, control logic, parameters, and structural connectivity. 
    & Helps connect tool messages, warnings, and reports back to the relevant modules, signals, parameters, or hierarchy in the original design. \\
    \midrule
    Synthesis script
    & \texttt{synthesis.tcl}
    & Specifies how the synthesis tool is invoked, including the top module, target library, constraints, optimization commands, and report generation steps.
    & Provides the run configuration needed to interpret logs, netlists, and PPA reports in the context of the synthesis procedure. \\
    \midrule
    Synthesis log 
    & \texttt{synthesis.log}
    & Records the synthesis run, including tool setup, loaded libraries, elaboration messages, warnings, errors, optimization steps, runtime, and final status. 
    & Serves as the main trace of what happened during synthesis and helps identify failures, unresolved references, missing constraints, and unexpected tool behavior. \\
    \midrule
    Netlist 
    & \texttt{netlist.v}
    & Represents the synthesized gate-level implementation produced from the RTL using a target technology library. It is typically consumed by downstream physical design, equivalence checking, and gate-level simulation. 
    & Allows analysis to compare intended RTL structure with the synthesized implementation, detect unresolved black boxes, and inspect technology-mapped design structure. \\
    \midrule
    PPA reports 
    & \makecell[c]{\texttt{area\_report.txt}\\\texttt{power\_report.txt}\\\texttt{timing\_report.txt}}
    & Summarize power, performance, and area outcomes after synthesis, including cell area, timing slack, critical paths, dynamic power, and leakage power. 
    & Provide quantitative evidence for design quality, enabling questions about timing closure, area growth, power behavior, and cross-library or cross-design comparisons. \\
    \bottomrule
    \end{tabular}
    \caption{Artifact types used in our EDA artifact dataset and their role in the chip design flow.}
    \label{tab:eda_artifacts}
\end{table*}

\subsection{Design Collection and Synthesis Artifact Generation}
\label{app:collection-validation}

This section expands on the four-phase collection stage described in the design collection. The process transforms public GitHub repositories into the curated design corpus $\mathcal{D} = \{d_r : r \in \mathcal{R}_2,\ v(r) = 1\}$ from which downstream EDA artifacts are generated. Here, $d_r$ denotes the design entry derived from repository $r$ after synthesis feasibility validation, so $\mathcal{D}$ contains validated designs rather than raw repositories. We provide detailed documentation of the filter logic, thresholds, and LLM evaluator configuration to ensure the methodology can be reproduced. Exact counts at each phase may differ across reruns because GitHub repository state (e.g., stars, archive status, and default-branch contents) evolves over time.

\paragraph{Table~\ref{tab:keywords}: discovery keyword set $\mathcal{K}$.}
The discovery phase queries the GitHub search API using $\mathcal{K}$, organized into three groups: 2 language filters (\texttt{language:Verilog}, \texttt{language:SystemVerilog}) that scope the search to HDL content; 13 topic keywords (e.g., \texttt{fpga}, \texttt{rtl}, \texttt{asic}, \texttt{risc-v}, \texttt{soc}) targeting design-style descriptors; and 33 design-specific keywords spanning processors, peripheral controllers, accelerators, and cryptographic blocks. The initial candidate set is $\mathcal{R}_0 = \bigcup_{k \in \mathcal{K}} \textsc{Search}(k)$ after URL-level deduplication. We intentionally made $\mathcal{K}$ broad to avoid biasing the dataset toward CPU-heavy repositories. The diverse keywords illustrated in Table~\ref{tab:keywords} reveal the wide range of hardware categories in Figure~\ref{fig:dataset_design}.

\paragraph{Table~\ref{tab:heuristic-rules}: heuristic predicate $h(r)$.}
Heuristic filtering removes repositories that are unlikely to contain usable HDL designs without requiring LLM intervention. A repository passes $h(r)$ if and only if every rule in Table~\ref{tab:heuristic-rules} holds: the repository is not archived; its primary language is one of Verilog, SystemVerilog, or VHDL; it has at least 2 stars, with a manually curated set of seed repositories exempt from this rule to retain canonical designs; its total size is at least 10\, KB; its most recent push is within ten years; non-trivial forks are retained only if the fork itself has at least five stars, filtering out passive fork copies; and neither the repository name nor description matches a case-insensitive regex flagging academic coursework ($\rho_{\mathrm{academic}}$). The retained set is $\mathcal{R}_1 = \{r \in \mathcal{R}_0 : h(r) = 1\}$.

\paragraph{Table~\ref{tab:gpt-evaluator}: LLM evaluator $\ell(r)$.}
Each $r \in \mathcal{R}_1$ is evaluated by \texttt{GPT-4.1-nano} with temperature $T = 0.1$ and response format constrained to JSON, using a fixed prompt template. The evaluator input consists of repository metadata, including name, star count, primary language, and description; a README prefix; and a filtered listing of HDL-relevant files. The accept criteria, reject criteria, and output schema are summarized in Table~\ref{tab:gpt-evaluator}. The acceptance rule is $\ell(r)=1$ iff the returned decision is \texttt{ACCEPT} and the returned confidence is at least $\tau_{\mathrm{conf}} = 0.6$, yielding $\mathcal{R}_2 = \{r \in \mathcal{R}_1 : \ell(r) = 1\}$. The confidence threshold reduces borderline accepts where the LLM is uncertain about synthesizability. Failed API calls or malformed JSON responses are treated as rejected decisions.

\paragraph{Table~\ref{tab:feasibility-check}: synthesis feasibility check $v(r)$.}
Although $\ell(r)$ rules out non-design repositories, it does not guarantee that a repository has the structural properties required by the synthesis process. The predicate $v(r)$ therefore checks for: (i) at least one non-testbench Verilog/SystemVerilog source file; (ii) at least one parseable \texttt{module ... endmodule} block recovered by regex-based parsing of comment-stripped source; (iii) at least one top-module candidate, defined as a module not instantiated by any other module, with all modules treated as candidates if every module is instantiated; and (iv) a recognizable clock input on the selected top module. Testbench files are excluded using both path components and filename patterns, as listed in Table~\ref{tab:feasibility-check}. Clock-port detection matches top-module input ports against one of 15 regexes, with a default \texttt{clk} fallback when no match is found. The retained corpus is $\mathcal{D} = \{d_r : r \in \mathcal{R}_2,\ v(r) = 1\}$.

\paragraph{Algorithm~\ref{alg:design_analysis}: static design analysis and top-module ranking.}
Algorithm~\ref{alg:design_analysis} formalizes the static analysis that $v(r)$ relies on and that phase~(ii) of the generation stage consumes. Given the repository's HDL file set $\mathcal{F}$, the procedure parses module declarations to extract names, port lists, and instantiations, and constructs the directed instantiation graph $G = (\mathcal{M}, E)$ with an edge $m_i \rightarrow m_j$ whenever $m_i$ instantiates $m_j$. The candidate root set $\mathcal{R}$ is the set of modules with in-degree zero in $G$. Each candidate $m \in \mathcal{R}$ is scored deterministically by the tuple $s_m = (\neg b_m,\, d_m,\, \mathrm{lines}(m),\, -\mathrm{name}(m))$, where $b_m$ flags testbench-like modules, $d_m$ is the size of the transitive instantiation footprint, and ties are broken by line count and lexicographic name. The ranked candidate list $\mathcal{T}$ is returned in descending order of $s_m$ and used by the iterative repair loop.

\paragraph{Filter funnel.}
From an initial $|\mathcal{R}_0| = 23{,}115$ discovered candidates, heuristic filtering retained $|\mathcal{R}_1| = 16{,}514$ ($71.4\%$); the LLM evaluator retained $|\mathcal{R}_2| = 8{,}812$ ($38.1\%$ of $\mathcal{R}_0$, $53.4\%$ of $\mathcal{R}_1$); and the synthesis feasibility check retained $|\mathcal{D}| = 2{,}787$ ($12.1\%$ of $\mathcal{R}_0$, $31.6\%$ of $\mathcal{R}_2$). The funnel illustrates that each gate is non-redundant: $h(r)$ primarily discards stale, non-HDL, or coursework repositories; $\ell(r)$ primarily discards semantically borderline content such as software projects, verification-only collections, and EDA tooling that heuristic rules cannot identify; and $v(r)$ primarily discards repositories whose module graphs are unbuildable, such as designs with no top-module candidate, no clock signal, or unparseable source.

The process is layered so that each filter addresses a different failure mode. $\mathcal{K}$ controls the breadth of design categories; $h(r)$ removes unmaintained or non-HDL repositories at near-zero cost; $\ell(r)$ provides semantic judgment about design intent that rules cannot easily express; and $v(r)$ enforces the structural preconditions required by the synthesis pipeline. The transitive-depth ranking from Algorithm~\ref{alg:design_analysis} is also reused within the iterative repair loop: when synthesis of the highest-ranked module fails, the next-ranked candidate is tried before declaring the design unsynthesizable. This deterministic ranking allows the repair loop to explore alternative tops without additional LLM calls and makes the bound $\min(15,\,2\,|\mathcal{T}|)$ a meaningful repair budget rather than an unconstrained hyperparameter.

\begin{table}[t]
    \centering
    \small
    \setlength{\tabcolsep}{1mm}
    \begin{tabular}{
    >{\raggedright\arraybackslash}p{0.22\columnwidth}
    >{\raggedright\arraybackslash}p{0.70\columnwidth}}
    \toprule
    \textbf{Category} & \textbf{Search Queries} \\
    \midrule
    Language (2)
    & \texttt{language:Verilog}, \texttt{language:SystemVerilog} \\
    \midrule
    Topic (13)
    & \texttt{fpga}, \texttt{rtl}, \texttt{asic}, \texttt{risc-v}, \texttt{riscv}, \texttt{processor}, \texttt{soc}, \texttt{ip-core}, \texttt{verilog}, \texttt{systemverilog}, \texttt{hdl}, \texttt{hardware-design}, \texttt{digital-design} \\
    \midrule
    Keyword (33)
    & \texttt{synthesizable verilog}, \texttt{AXI}, \texttt{UART verilog}, \texttt{SPI verilog}, \texttt{I2C verilog}, \texttt{PCIe verilog}, \texttt{DDR controller}, \texttt{FIFO verilog}, \texttt{NoC router}, \texttt{cache controller verilog}, \texttt{ethernet verilog}, \texttt{USB verilog}, \texttt{JTAG verilog}, \texttt{DMA controller}, \texttt{interrupt controller}, \texttt{RISC-V core}, \texttt{ARM processor verilog}, \texttt{MIPS processor verilog}, \texttt{GPU verilog}, \texttt{DSP verilog}, \texttt{FPU verilog}, \texttt{crypto verilog AES}, \texttt{crypto verilog SHA}, \texttt{neural network accelerator verilog}, \texttt{CORDIC verilog}, \texttt{CRC verilog}, \texttt{PWM verilog}, \texttt{ADC interface verilog}, \texttt{memory controller verilog}, \texttt{bus arbiter verilog}, \texttt{wishbone}, \texttt{APB verilog} \\
    \bottomrule
    \end{tabular}
    \caption{Keywords $\mathcal{K}$ used in Discovery phase.}
    \label{tab:keywords}
\end{table}

\begin{table}[t]
    \centering
    \small
    \setlength{\tabcolsep}{1mm}
    \begin{tabular}{
    >{\centering\arraybackslash}p{0.16\columnwidth}
    >{\raggedright\arraybackslash}p{0.38\columnwidth}
    >{\raggedright\arraybackslash}p{0.34\columnwidth}}
    \toprule
    \textbf{No.} & \textbf{Rule} & \textbf{Threshold} \\
    \midrule
    (i) & Repository not archived & $r.\text{archived}=\text{False}$ \\
    (ii) & HDL primary language & $\in \{\text{Verilog}, \text{SystemVerilog}\}$ \\
    (iii) & Minimum stars & $\geq 2$; seed repositories exempt \\
    (iv) & Minimum repository size & $\geq 10$~KB \\
    (v) & Activity recency & last push $\leq 10$ years \\
    (vi) & Non-trivial fork & if fork: stars $\geq 5$ \\
    (vii) & Not academic coursework & regex $\rho_{\text{academic}}$ on name $\cup$ description \\
    \bottomrule
    \end{tabular}
    \caption{Heuristic rules constituting $h(r)$.}
    \label{tab:heuristic-rules}
\end{table}

\begin{table}[t]
    \centering
    \small
    \setlength{\tabcolsep}{1mm}
    \begin{tabular}{p{0.28\linewidth} p{0.62\linewidth}}
    \toprule
    \textbf{Item} & \textbf{Description} \\
    \midrule
    Model & GPT-4.1-nano, temperature $T=0.1$ \\
    \midrule
    Input fields & Repository name, stars, language, description, README prefix, filtered HDL-relevant file tree \\
    \midrule
    Accept criteria & Synthesizable RTL, IP cores, processors, SoCs, memory controllers, DSP/FPU blocks, crypto accelerators, communication interfaces, FPGA designs, hardware security modules \\
    \midrule
    Reject criteria & Coursework, simulation-only testbenches, EDA scripts/tools, trivial designs, unmodified forks, documentation-only repositories, software projects, HLS source, board support packages, verification-only repositories with no DUT \\
    \midrule
    Output schema & JSON object with \texttt{decision}, \texttt{confidence}, \texttt{category}, and \texttt{reasoning} \\
    \midrule
    Acceptance rule & $\ell(r)=1$ iff \texttt{decision}=\texttt{ACCEPT} and \texttt{confidence} $\geq \tau_{\text{conf}}=0.6$ \\
    \bottomrule
    \end{tabular}
    \caption{LLM evaluation $\ell(r)$ for design collection.}
    \label{tab:gpt-evaluator}
\end{table}

\begin{table}[t]
    \centering
    \small
    \setlength{\tabcolsep}{1mm}
    \begin{tabular}{p{0.34\linewidth} p{0.56\linewidth}}
    \toprule
    \textbf{Criterion} & \textbf{Configuration} \\
    \midrule
    Source file presence & At least one Verilog/SystemVerilog file (\texttt{.v},
    \texttt{.sv}) after testbench exclusion \\
    \midrule
    Testbench exclusion & Drop files whose path contains any of \{\texttt{tb},
    \texttt{test}, \texttt{tests}, \texttt{sim}, \texttt{verification},
    \texttt{uvm}, \texttt{cocotb}, \texttt{testbench}, \texttt{bench}\}, or
    whose filename matches \texttt{*\_tb.s?v}, \texttt{tb\_*.s?v},
    \texttt{*\_test.s?v}, \texttt{*\_testbench.s?v} \\
    \midrule
    Parseable modules & At least one \texttt{module $\dots$ endmodule} block
    recovered by regex-based parsing of comment-stripped source \\
    \midrule
    Top-module candidate & At least one module not instantiated by any other;
    candidates ranked by transitive instantiation depth (descending). If all
    modules are instantiated, every module becomes a candidate \\
    \midrule
    Clock-port detection & Top-module input port matching one of 15 regexes:
    \texttt{clk}, \texttt{clock}, \texttt{i\_clk}, \texttt{i\_clock},
    \texttt{sys\_clk}, \texttt{aclk}, \texttt{pclk}, \texttt{hclk},
    \texttt{mclk}, \texttt{core\_clk}, \texttt{clk\_i}, \texttt{wb\_clk\_i},
    \texttt{wbclk}, \texttt{*\_clk}, \texttt{clk\_*} (fallback to default
    \texttt{clk} if none match) \\
    \bottomrule
    \end{tabular}
    \caption{Synthesis feasibility check $v(r)$ used to validate collected repositories.}
    \label{tab:feasibility-check}
\end{table}

Algorithm~\ref{alg:design_analysis} describes the static design analysis procedure used to rank candidate top modules. The iterative repair loop then applies deterministic fixes, including top-module changes, search-path extension, and clock selection, until synthesis succeeds, no useful candidates remain, or the attempt threshold is reached.


\begin{algorithm}[t]
\small
\caption{Static Design Analysis and Top-Module Ranking}
\label{alg:design_analysis}
\begin{algorithmic}[1]
\Require Repository files $\mathcal{F}$
\Ensure Ranked top-module candidates $\mathcal{T}$

\State $\mathcal{V} \gets$ Verilog/SystemVerilog files in $\mathcal{F}$
\State $\mathcal{M} \gets \textsc{ParseModules}(\mathcal{V})$
\Statex \hspace{\algorithmicindent}{ Inputs: name, file, ports, lines, instantiations}

\State Construct instantiation graph $H=(\mathcal{M},E)$
\Statex \hspace{\algorithmicindent}{ Edge $m_i \rightarrow m_j$ if $m_i$ instantiates $m_j$}

\State $\mathcal{R} \gets \{m \in \mathcal{M}: \mathrm{indegree}_G(m)=0\}$
\Statex \hspace{\algorithmicindent}{ Candidate root modules}

\ForAll{$m \in \mathcal{R}$}
    \State $d_m \gets \textsc{ReachableSize}(H,m)$
    \State $b_m \gets \textsc{IsTestbenchLike}(m)$
    \State $s_m \gets (\neg b_m,\ d_m,\ \mathrm{lines}(m),\ -\mathrm{name}(m))$
\EndFor

\State $\mathcal{T} \gets$ modules in $\mathcal{R}$ sorted by $s_m$ descending
\State \Return $\mathcal{T}$
\end{algorithmic}
\end{algorithm}

\subsection{Knowledge Graph Construction}
\label{app:kg-construction}

This section explains the knowledge graph construction process. \ourtool{} represents each design as a typed graph with nodes for designs, artifacts, modules, warnings, reports, metrics, and tool stages. Edges encode artifact membership, chunk containment, generation provenance, module definitions, reported metrics, and message references.

Algorithm 3 presents the concrete ingestion procedure for instantiating the offline representation introduced in the offline stage section. For each design \(d\), the algorithm first creates a design-level anchor node and then partitions the ingestion set \(I_d\) based on the parser to be applied. Textual artifacts, including documentation, logs, reports, and synthesis scripts, are inserted as artifact nodes and then divided into overlapping chunks. Each chunk node \(v_c^{(k)}\) stores graph metadata such as its source artifact and span, while its embedding \(z_c^{(k)}\) is stored separately in the vector index as \((v_c^{(k)}, z_c^{(k)})\). This allows semantic retrieval to return compact text evidence while preserving a direct path back to the original design and artifact context in the graph.

Structural artifacts, including RTL and netlist files, are handled differently. Rather than being represented only as text chunks, they are parsed into syntax-level subgraphs whose nodes and edges capture the artifact's internal structure, such as modules, instances, signals, or netlist elements. These structural nodes are attached to their corresponding artifact nodes via containment edges. The final output is therefore a global graph \(G=(V,E)\) and vector index \(Z\), where text evidence can be retrieved through \(Z\) and then resolved through \(G\) to recover artifact hierarchy, design provenance, and structural context.


\begin{algorithm}[!t]
\footnotesize
\caption{Knowledge Graph Ingestion}
\label{alg:eda_artifact_ingestion}
\begingroup
\newcommand{\AlgI}{\hspace{1.4em}}
\newcommand{\AlgII}{\hspace{2.8em}}
\newcommand{\AlgIII}{\hspace{4.2em}}
\begin{algorithmic}[1]
\Require Design dataset $\mathcal{D}$, where each $d$ has ingestion set $\mathcal{I}_d$; chunk size $W$, overlap $\Omega$
\Ensure Attributed typed knowledge graph $G=(V,E)$ and graph-associated vector index $\mathcal{Z}$

\State $G \gets (V,E) \gets (\emptyset,\emptyset)$; $\mathcal{Z} \gets \emptyset$

\State \textbf{for each} design $d \in \mathcal{D}$ \textbf{do}
    \State \AlgI Create design node $v_d$ with metadata $\phi(v_d)$
    \State \AlgI $V \gets V \cup \{v_d\}$

    \Statex
    \State \AlgI \Comment{Artifact parsing}
    \State \AlgI $\mathcal{I}_d^{\mathrm{text}} \gets$ documents, logs, reports, and scripts in $\mathcal{I}_d$
    \State \AlgI $\mathcal{I}_d^{\mathrm{struct}} \gets$ RTL and netlist files in $\mathcal{I}_d$

    \Statex
    \State \AlgI \Comment{Semantic indexing}
    \State \AlgI \textbf{for each} text artifact $a \in \mathcal{I}_d^{\mathrm{text}}$ \textbf{do}
        \State \AlgII Create artifact node $v_a$ with metadata $\phi(v_a)$
        \State \AlgII $V \gets V \cup \{v_a\}$
        \State \AlgII $E \gets E \cup \{(v_d,\mathrm{HAS\_ARTIFACT},v_a)\}$

        \State \AlgII $n_a \gets \left\lceil \frac{L(a)-\Omega}{W-\Omega} \right\rceil$
        \State \AlgII \textbf{for} $n=1$ \textbf{to} $n_a$ \textbf{do}
            \State \AlgIII $\ell_s^{(n)} \gets 1+(n-1)(W-\Omega)$
            \State \AlgIII $\ell_e^{(n)} \gets \min(\ell_s^{(n)}+W-1, L(a))$
            \State \AlgIII Create chunk node $v_c^{(n)}$ with metadata $\phi(v_c^{(n)})$
            \State \AlgIII $z_c^{(n)} \gets \textsc{Embed}(v_c^{(n)})$
            \State \AlgIII $\mathcal{Z} \gets \mathcal{Z} \cup \{(v_c^{(n)},z_c^{(n)})\}$
            \State \AlgIII $V \gets V \cup \{v_c^{(n)}\}$
            \State \AlgIII $E \gets E \cup \{(v_a,\mathrm{HAS\_CHUNK},v_c^{(n)})\}$
        \State \AlgII \textbf{end for}
    \State \AlgI \textbf{end for}

    \Statex
    \State \AlgI \Comment{Structural parsing}
    \State \AlgI \textbf{for each} structural artifact $s \in \mathcal{I}_d^{\mathrm{struct}}$ \textbf{do}
        \State \AlgII Create artifact node $v_s$ with metadata $\phi(v_s)$
        \State \AlgII $V \gets V \cup \{v_s\}$
        \State \AlgII $E \gets E \cup \{(v_d,\mathrm{HAS\_ARTIFACT},v_s)\}$

        \State \AlgII $T_s \gets \textsc{ParseAST}(s)$
        \State \AlgII $(V_s,E_s) \gets \textsc{BuildSyntaxGraph}(T_s)$
        \State \AlgII $V \gets V \cup V_s$
        \State \AlgII $E \gets E \cup E_s$
        \State \AlgII $E \gets E \cup \{(v_s,\mathrm{CONTAINS},v): v \in V_s\}$
    \State \AlgI \textbf{end for}
\State \textbf{end for}

\State \Return $G=(V,E), \mathcal{Z}$
\end{algorithmic}
\endgroup
\end{algorithm}

\subsection{Creation of EDA Evaluation Benchmark}
\label{app:evaluation-scoring}

This section describes the creation of the EDA artifact analysis QA benchmark, which is used to evaluate \ourtool{} against commercial agentic frameworks in the results section. The benchmark contains $90$ question-answer (QA) pairs organized along 2 axes: 3 question types (factual, statistical, and reasoning) and 3 difficulty levels (easy, medium, and hard), with $10$ QA pairs in each $(\text{type}, \text{level})$ cell. Table~\ref{tab:eval_benchmark} lists the criteria used to assign each level. The 3 question types evaluate deep analytical capabilities: factual questions test retrieval of explicit information from artifacts; statistical questions test retrieval combined with aggregation or numerical computation; and reasoning questions test grounded cross-artifact reasoning across logs, netlists, source files, reports, and other artifacts.

The initial QA drafts were generated using \texttt{Gemini~3.1~Pro}, prompted with samples drawn from the curated EDA artifact dataset $\mathcal{D}$. For each draft, the model generated a candidate question based on one or more artifacts from a specific design, along with a corresponding candidate answer. Using a large generative model instead of manually writing every question enabled a wider range of artifact types and phrasing styles than would have been possible with the same effort budget. However, the generated pairs were treated as initial drafts, with all final benchmark content determined by the authors, as described next.

Each candidate QA pair was reviewed and modified by the authors using the actual EDA artifact dataset before being included in the benchmark. The authors performed three functions for each pair: (i) \textbf{grounding}: questions were rewritten to reference specific designs and artifacts in $\mathcal{D}$, and any pair that could not be grounded was discarded; (ii) \textbf{golden-answer verification}: answers were manually re-extracted by inspecting the cited artifacts, ensuring numeric values were sourced from PPA reports, log evidence was located in the corresponding synthesis log, and structural claims were verified against the netlist or source code until they matched the referenced artifacts; and (iii) \textbf{difficulty assignment}: the authors assigned a $(\text{type},\,\text{level})$ label based on the modified question and its supporting evidence requirements, according to the criteria in Table~\ref{tab:eval_benchmark}, rather than carrying over any label suggested by \texttt{Gemini~3.1~Pro}. This protocol ensures that the final benchmark content and labels are author-verified and that all QA pairs are grounded in checkable artifacts from $\mathcal{D}$.

\begin{table}[t]
\centering
\small
\begin{tabular}{lll}
\toprule
\textbf{Type} & \textbf{Level} & \textbf{Criteria} \\
\midrule
\multirow{3}{*}{Factual}
  & Easy   & Single-file lookup \\
  & Medium & Multi-file look-up \\
  & Hard   & Multi-step extraction \\
\midrule
\multirow{3}{*}{Statistical}
  & Easy   & Basic arithmetic / comparison \\
  & Medium & Multiple statistical measures \\
  & Hard   & Extracted trend / relationships \\
\midrule
\multirow{3}{*}{Reasoning}
  & Easy   & Simple interpretation \\
  & Medium & Connecting 2 / 3 observations \\
  & Hard   & Deep domain expertise \\
\bottomrule
\end{tabular}
\caption{Evaluation benchmark question taxonomy.}
\label{tab:eval_benchmark}
\end{table}

\subsection{Human Expert Benchmark Evaluation Process}
To then evaluate the agent-generated results of this benchmark, we utilize manual scoring by human experts (as described in the experiments section). Table~\ref{tab:scoring_rubric} provides the manual scoring rubric, detailing the intuition provided to reviewers behind the scoring scale (1-10) when assessing the agent response in relation to the golden solution. To mitigate evaluation bias and ensure fair evaluation, the reviewers are provided the responses from each agent/model configuration in a randomized order for each benchmark question, ensuring blindness to the model and agent origin. 

\subsection{LLM Benchmark Evaluation Process}
To supplement the human expert evaluations, we additionally utilize the LLM-council evaluation method to score the agent responses. Within this process, two judge models (\texttt{GPT-5.4} and \texttt{Claude Sonnet-4.6}) independently score each answer against the golden answer according to the same evaluation rubric~\ref{tab:scoring_rubric}. Then, a chairman model (\texttt{Claude Opus-4.7}) is provided with the context of both evaluations and determines the final score.

\begin{table}[t]
    \centering
    \small
    \setlength{\tabcolsep}{1mm}
    \begin{tabular}{ c >{\centering\arraybackslash}p{0.72\columnwidth} }
    \toprule
    \textbf{Score} & \textbf{Meaning} \\
    \midrule
    10 & Excellent; fully correct; no errors. \\
    \midrule
    8 & Good; mostly correct, with minor gaps or imprecision. \\
    \midrule
    6 & Partial; captures part of the answer, but has significant gaps. \\
    \midrule
    4 & Weak; touches on the topic but is mostly wrong or missing key elements. \\
    \midrule
    2 & Wrong answer, but no hallucination. \\
    \midrule
    0 & Wrong answer with hallucination. \\
    \bottomrule
    \end{tabular}
    \caption{Scoring rubric for human expert evaluation.}
    \label{tab:scoring_rubric}
\end{table}

\begin{figure*}[t]
    \centering

    \begin{subfigure}{\linewidth}
        \centering
        \includegraphics[width=\linewidth]{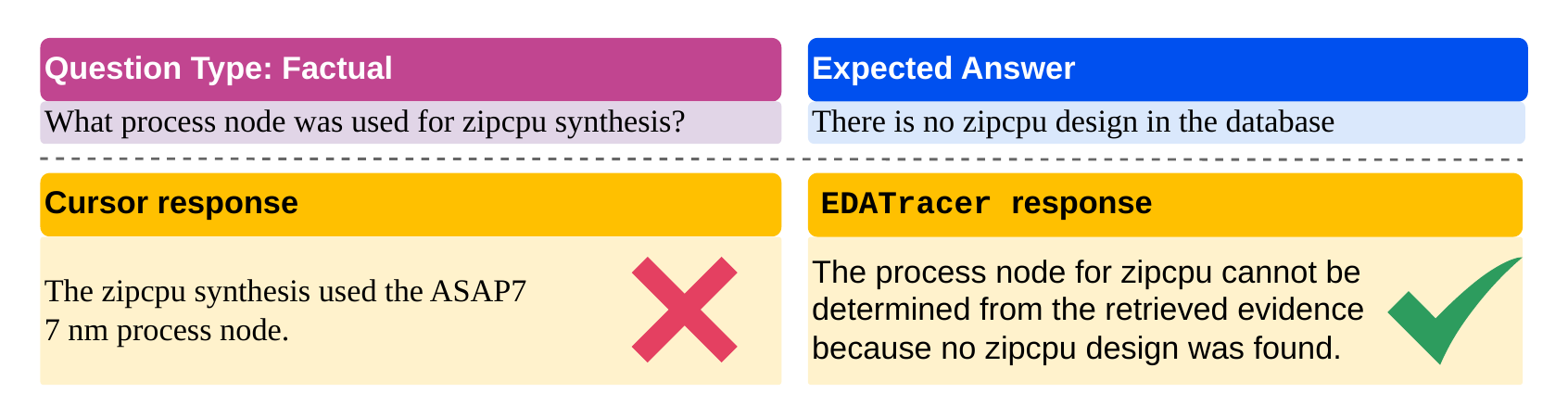}
        \caption{Factual question-answer comparison of \ourtool{} v/s Cursor CLI.}
        \label{fig:ex1}
    \end{subfigure}

    \begin{subfigure}{\linewidth}
        \centering
        \includegraphics[width=\linewidth]{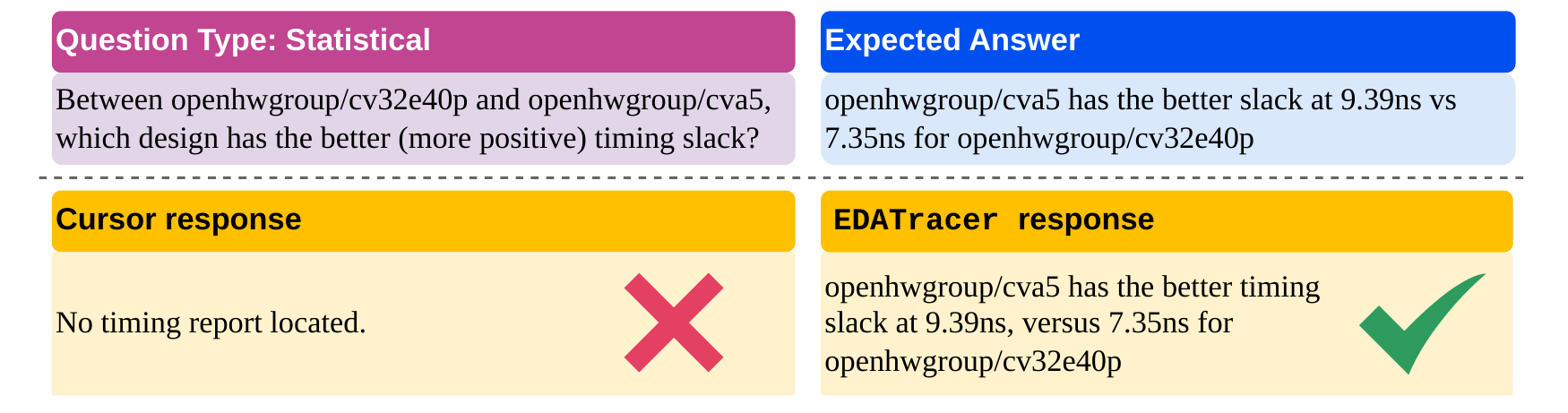}
        \caption{Statistical question-answer comparison of \ourtool{} v/s Cursor CLI.}
        \label{fig:ex2}
    \end{subfigure}

    \begin{subfigure}{\linewidth}
        \centering
        \includegraphics[width=\linewidth]{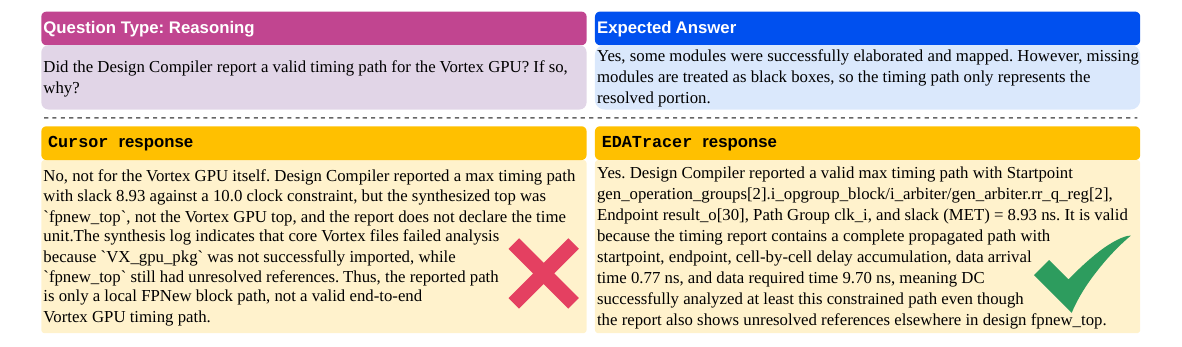}
        \caption{Reasoning question-answer comparison of \ourtool{} v/s Cursor CLI.}
        \label{fig:ex3}
    \end{subfigure}

    \caption{Question-answer comparison of \ourtool{} v/s Cursor CLI across factual, statistical, and reasoning questions.}
    \label{fig:eda-tool-comparison}
\end{figure*}

\subsection{Qualitative Examples}
\label{app:additional-results}

This section presents 3 QA examples drawn from the QA benchmark (defined in Section~\ref{app:evaluation-scoring}), comparing \ourtool{} against Cursor on 1 question from each category: factual, statistical, and reasoning. The examples illustrate 3 distinct failure modes that arise when retrieval is insufficiently grounded in the design-specific artifact structure.

Figure~\ref{fig:ex1} shows a factual question about the process node used for \texttt{zipcpu} synthesis. The correct answer is that the \texttt{zipcpu} design does not exist in the EDA artifact dataset. \ourtool{} returns this absence of design, whereas Cursor states that \texttt{zipcpu} used the ASAP7 7\,nm process node. This example illustrates a grounding failure. When design-specific evidence is missing, the correct behavior is to say that the answer cannot be determined. \ourtool{} avoids this failure by first resolving the queried design and grounding the answer based on the retrieved artifacts for that design.

Figure~\ref{fig:ex2} shows a statistical question comparing timing slack between \texttt{openhwgroup/cv32e40p} and \texttt{openhwgroup/cva5}. The correct answer reports raw slack values of $7.35$ and $9.39$, with \texttt{cva5} having the more positive slack. \ourtool{} retrieves the relevant timing reports and returns the comparison. Cursor fails to locate the timing report evidence, even though the artifacts are present in the dataset. This example illustrates a retrieval miss: the requested information exists, but the framework fails to surface the correct artifacts for the specified designs. \ourtool{} addresses this by resolving each design and retrieving the corresponding timing artifacts through the structured artifact representation.

Figure~\ref{fig:ex3} shows a reasoning question asking whether Design Compiler reported a valid timing path for the Vortex GPU design. The correct answer is nuanced: Design Compiler reports a max timing path with slack $\mathrm{MET}=8.93$, but the report must be interpreted in the context of unresolved references and the synthesized scope. \ourtool{} captures both parts of the evidence: it identifies the reported timing path and also notes the limitations implied by unresolved design references. Cursor instead treats the unresolved references as sufficient to reject the reported path. This example illustrates a failure of reasoning in which real evidence is retrieved but weighted incorrectly. \ourtool{} handles the case better because it considers the timing report and synthesis-log evidence together rather than allowing one artifact to override the other.

Thus, the three examples demonstrate that the key difference between \ourtool{} and Cursor is not merely whether relevant text is retrieved. The more challenging issues are whether the retrieved evidence aligns with the correct design, whether missing evidence is acknowledged, and whether multiple artifacts are interpreted in context with one another. In the factual example, we see how hallucination can occur due to missing evidence. The statistical example illustrates a retrieval miss, while the reasoning example reveals misjudged cross-artifact evidence. These are the precise instances where a typed, design-aware artifact representation proves beneficial: it provides the agent with a clear design reference, connects each design to its associated artifacts, and facilitates evidence collection across the different kinds of artifacts.

\begin{table*}[t]
    \centering
    \setlength{\tabcolsep}{1mm}
    \small
    \begin{tabular}{llccccccccc}
        \toprule
        \multirow{2}{*}{Model} & \multirow{2}{*}{Tool}
        & \multicolumn{4}{c}{Human Expert Average Score}
        & \multicolumn{4}{c}{LLM Council Score}
        & \multirow{2}{*}{Token Usage} \\
        \cmidrule(lr){3-6} \cmidrule(lr){7-10}
        & & Factual & Statistical & Reasoning & Overall
          & Factual & Statistical & Reasoning & Overall
          & \\
        \midrule

        \multirow{2}{*}{Claude-Haiku-4.5}
        & RAG  & 3.37 & 2.00 & 2.37 & 2.58 & 4.50 & 3.27 & 3.70 & 3.82 & \bc 24.9M \\
        & KG & \bc 8.83   & \bc 9.53   & \bc 7.67   & \bc 8.68   & \bc 8.73   & \bc 9.43   & \bc 7.17   & \bc 8.44 & 36.4M \\
        \midrule
        
        \multirow{2}{*}{GPT-5.4-mini}
        & RAG     & 3.20   & 0.20   & 1.0   & 1.47   & 4.53 & 3.00 & 2.60 & 3.19 & \bc 14.5M \\
        & KG & \bc 8.07   & \bc 7.30   & \bc 6.83   & \bc 7.40   & \bc 8.80 & \bc 7.67 & \bc 7.33 & \bc 7.93 & 19.9M \\

        \bottomrule
    \end{tabular}
    \caption{Comparison between \ourtool{} using a RAG vs KG}
    \label{tab:kg_rag_ablation}
\end{table*}

\begin{table*}[t]
    \centering
    \small
    \setlength{\tabcolsep}{1mm}
    \begin{tabular}{ll*{12}{c}c}
        \toprule
        \multirow{2}{*}{\textbf{Model}}
        & \multirow{2}{*}{\textbf{Tool}}
        & \multicolumn{3}{c}{\textbf{Factual}}
        & \multicolumn{3}{c}{\textbf{Statistical}}
        & \multicolumn{3}{c}{\textbf{Reasoning}}
        & \multicolumn{3}{c}{\textbf{Overall}}
        & \multirow{2}{*}{\textbf{Tokens}} \\
        \cmidrule(lr){3-5}
        \cmidrule(lr){6-8}
        \cmidrule(lr){9-11}
        \cmidrule(lr){12-14}
        & & \textbf{Avg.} & \textbf{P@1} & \textbf{P@5}
        & \textbf{Avg.} & \textbf{P@1} & \textbf{P@5}
        & \textbf{Avg.} & \textbf{P@1} & \textbf{P@5}
        & \textbf{Avg.} & \textbf{P@1} & \textbf{P@5}
        & \\
        \midrule

        \multirow{3}{*}{Claude Opus 4.7}
        & Cursor      & 9.60 & 96.0 & \bc 100 & 9.53 & 94.0 & \bc 100 & \bc 8.40 & \bc 76.7 & 96.7 & 9.18 & 88.9 & 98.9 & 78.4M \\
        & Claude Code & 9.13 & 74.7 & \bc 100 & 9.40 & 82.0 & \bc 100 & 7.80 & 60.7 & \bc 100 & 8.78 & 72.4 & \bc 100 & 62.7M \\
        & \ourtool{}  & \bc 9.83 & \bc 99.3 & \bc 100 & \bc 9.57 & \bc 97.3 & \bc 100 & 8.33 & 76.0 & 86.7 & \bc 9.24 & \bc 90.9 & 95.6 & \bc 30.0M \\
        \midrule

        \multirow{3}{*}{Claude Sonnet 4.6}
        & Cursor      & 9.20 & 92.0 & \bc 100 & 9.40 & 90.0 & \bc 100 & \bc 8.07 & 72.0 & \bc 90.0 & 8.89 & 84.7 & \bc 96.7 & 67.8M \\
        & Claude Code & \bc 9.43 & \bc 94.7 & 96.7 & 9.33 & \bc 96.7 & \bc 100 & 7.97 & 77.3 & \bc 90.0 & 8.91 & \bc 89.6 & 95.6 & 64.0M \\
        & \ourtool{}  & 9.20 & 92.0 & 96.7 & \bc 9.57 & 96.0 & \bc 100 & 8.03 & \bc 78.7 & \bc 90.0 & \bc 8.93 & 88.9 & 95.6 & \bc 30.0M \\
        \midrule

        \multirow{3}{*}{Claude Haiku 4.5}
        & Cursor      & \bc 9.60 & \bc 96.7 & \bc 100.0 & 8.47 & 90.7 & \bc 100 & \bc 7.33 & \bc 61.3 & 76.7 & 8.47 & 77.8 & \bc 92.2 & 84.6M \\
        & Claude Code & 9.57 & 92.0 & 96.7 & 9.07 & 91.3 & 96.7 & \bc 7.33 & 60.7 & \bc 83.3 & \bc 8.66 & \bc 81.3 & \bc 92.2 & 115M \\
        & \ourtool{}  & 8.73 & 86.0 & 93.3 & \bc 9.43 & \bc 94.7 & \bc 100 & 7.17 & 59.3 & 76.7 & 8.44 & 80.0 & 90.0 & \bc 36.4M \\
        \midrule

        \multirow{2}{*}{GPT 5.4}
        & Cursor      & \bc 9.23 & \bc 94.7 & \bc 96.7 & 8.43 & 86.0 & 96.7 & 6.90 & 64.7 & 83.3 & 8.19 & 81.8 & 92.2 & 53.6M \\
        & \ourtool{}  & 8.97 & 92.0 & \bc 96.7 & \bc 9.30 & \bc 92.7 & \bc 100 & \bc 8.90 & \bc 76.7 & \bc 9.06 & \bc 96.7 & \bc 90.0 & \bc 97.8 & \bc 26.2M \\
        \midrule

        \multirow{2}{*}{GPT 5.4 mini}
        & Cursor      & \bc 8.80 & \bc 88.7 & \bc 96.7 & 7.67 & \bc 83.3 & \bc 100 & \bc 7.33 & \bc 58.7 & \bc 90.0 & \bc 7.93 & \bc 77.8 & \bc 95.6 & 59.3M \\
        & \ourtool{}  & 8.00 & 67.3 & 86.7 & \bc 7.97 & 79.3 & 86.7 & 6.77 & 52.0 & 83.3 & 7.58 & 71.1 & 85.6 & \bc 19.9M \\
        \midrule

        \multirow{1}{*}{Qwen 3.5 9B}
        & \ourtool{}  & 8.03 & 74.0 & 90.0 & 8.37  & 85.3 & 100 & 6.07 & 36.0 & 73.3 & 7.49 & 65.1 & 87.8 &  64.1M \\

        \bottomrule
    \end{tabular}%
    \caption{Accuracy comparison between \ourtool{} and other agents using LLM-council scoring.}
    \label{tab:comparison_accuracy_llm}
\end{table*}

\subsection{Ablation Studies}
\subsubsection{Database Formulation}
\label{app:ablation-database}

A central design decision in \ourtool{} is to structure EDA artifacts into a typed knowledge graph (KG) rather than a flat vector store used by standard retrieval-augmented generation (RAG). This ablation isolates the contribution of that choice by comparing the same underlying LLM against two retrieval methods: (i) a RAG database built over the same EDA artifact dataset, and (ii) \ourtool{}'s KG described in the offline stage. The QA evaluation benchmark, scoring rubric, and prompts are held fixed; only the retrieval method changes. Results are reported in Table~\ref{tab:kg_rag_ablation}.

The RAG baseline chunks every artifact in $\mathcal{D}$ using the same chunk size $W$ and overlap $\Omega$ as those used by the KG semantic index. It also embeds the chunks using the same embedding model and retrieves the top-$K$ chunks based on semantic similarity. Unlike the KG formulation, this baseline does not include typed nodes, artifact provenance, design hierarchy, or graph neighborhoods. The comparison, therefore, tests whether structured artifact relationships improve EDA artifact analysis beyond flat semantic retrieval.

Table~\ref{tab:kg_rag_ablation} shows that the KG formulation substantially improves accuracy across both LLM backbones. With \texttt{Claude Haiku 4.5}, the human expert overall score increases from $2.58$ with RAG to $8.68$ with KG, a $+6.10$ absolute gain and a \textbf{3.36$\times$} improvement. The LLM-council overall score increases from $3.82$ to $8.44$, a $+4.62$ absolute gain and a \textbf{2.21$\times$} improvement. With \texttt{GPT-5.4-mini}, the human expert overall score increases from $1.47$ to $7.40$, a $+5.93$ absolute gain and a \textbf{5.03$\times$} improvement, while the LLM-council overall score increases from $3.19$ to $7.93$, a $+4.74$ absolute gain and a \textbf{2.49$\times$} improvement. These gains appear across factual, statistical, and reasoning categories, with especially large improvements on statistical and reasoning questions where the system must connect PPA reports, logs, source files, and generated outputs.

The KG formulation uses more tokens than the RAG baseline: $36.4$M vs.\ $24.9$M for \texttt{Claude Haiku 4.5}, an increase of $11.5$M tokens or \textbf{1.46$\times$}, and $19.9$M vs.\ $14.5$M for \texttt{GPT-5.4-mini}, an increase of $5.4$M tokens or \textbf{1.37$\times$}. This increase is expected because the KG database includes graph-resolved context, such as artifact metadata and neighboring evidence, rather than only the retrieved text chunks. However, the token increase is modest compared with the accuracy gain. Therefore, the results suggest that the added structured context is not solely extra prompt content; it provides useful grounding for cross-artifact analysis.

Thus, the above ablation study supports the main design choice behind \ourtool{}. Flat RAG can retrieve locally relevant text, but it lacks an explicit mechanism to preserve which design, synthesis run, artifact type, or generated output a chunk belongs to. As a result, it can retrieve plausible but poorly scoped evidence. The KG formulation keeps retrieved evidence connected to its design context and provenance, allowing the agent to reason over related artifacts rather than isolated chunks. This advantage is especially important for smaller models, where the retrieval substrate plays a larger role in constraining the answer space.

\subsubsection{Benchmark Scoring Mechanism}
Here, we detail how the human expert evaluations of the EDA-evaluation benchmark described previously in the results compare against the LLM-Council evaluations across all agent configurations. Moreover, we provide the full LLM-Council results (average score and pass@1/@5)  in Table~\ref{tab:comparison_accuracy_llm} for the same set of agent responses. Additionally, we depict the net differences in each metric between the two strategies in Table~\ref{tab:comparison_accuracy_absdiff}.

Overall, we find the following similarities between the LLM-Council and human expert scoring. \ourtool{} resulted in the highest average score in relation to other agents in the three large LLMs (Opus 4,7, Sonnet 4.6, and GPT-5.4) across both evaluation methodologies. However, we find that for smaller models (e.g., Haiku 4.5 and 5.4 mini), Cursor and/or Claude Code averages exceed \ourtool{} 's LLM-Council results, in contrast to the human expert scoring for the Haiku 4.5 model. 

Furthermore, Table~\ref{tab:comparison_accuracy_absdiff} demonstrates that LLM-Council frequently results in a net increase in average scores relative to human expert scoring, primarily in Cursor and Claude Code agents, demonstrating a positive score increase between (+0.10-1.36) out of 10 across all underlying LLMs, with the exception of Cursor with Claude Sonnet 4.6. In contrast, LLM-council average scores remain consistent or degrade for \ourtool{}, with net changes ranging between (-0.3 to +0.18). These trends in average score shifts also impact the downstream pass@1 and pass@5 rates, given the threshold for a ``success'' remains consistent at 8. For instance, in pass@1, we see \ourtool{} outperforms Claude Code and Cursor in 4 of 5 LLMs with human expert scoring, and 2 of 5 LLMs with LLM-Council. This difference highlights the impact of the threshold score used to delineate successes and failures.

\definecolor{DiffPurple}{RGB}{111,78,161}

\newcommand{\absdiff}[2]{\cellcolor{DiffPurple!#1}\textcolor{black}{\textbf{#2}}}
\newcommand{\zerodiff}[1]{\cellcolor{black!5}\textcolor{black}{#1}}

\begin{table*}[t]
    \centering
    \small
    \setlength{\tabcolsep}{1mm}
    \begin{tabular}{ll*{12}{c}}
        \toprule
        \multirow{3}{*}{\textbf{Model}} 
        & \multirow{3}{*}{\textbf{Tool}} 
        & \multicolumn{3}{c}{\textbf{Factual}} 
        & \multicolumn{3}{c}{\textbf{Statistical}} 
        & \multicolumn{3}{c}{\textbf{Reasoning}} 
        & \multicolumn{3}{c}{\textbf{Overall}} \\
        \cmidrule(lr){3-5} 
        \cmidrule(lr){6-8} 
        \cmidrule(lr){9-11} 
        \cmidrule(lr){12-14}
        & 
        & \textbf{Avg.} & \textbf{pass@1} & \textbf{pass@5}
        & \textbf{Avg.} & \textbf{pass@1} & \textbf{pass@5}
        & \textbf{Avg.} & \textbf{pass@1} & \textbf{pass@5}
        & \textbf{Avg.} & \textbf{pass@1} & \textbf{pass@5} \\
        &
        & \textbf{(/10)} & \textbf{(pp)} & \textbf{(pp)}
        & \textbf{(/10)} & \textbf{(pp)} & \textbf{(pp)}
        & \textbf{(/10)} & \textbf{(pp)} & \textbf{(pp)}
        & \textbf{(/10)} & \textbf{(pp)} & \textbf{(pp)} \\
        \midrule

        \multirow{3}{*}{Claude Opus 4.7}
        & Cursor      & \absdiff{11}{0.20} & \absdiff{17}{9.3} & \absdiff{18}{10.0} & \absdiff{27}{1.20} & \absdiff{35}{27.3} & \absdiff{25}{16.7} & \absdiff{16}{0.47} & \absdiff{11}{2.7} & \zerodiff{0.0} & \absdiff{18}{0.62} & \absdiff{21}{13.1} & \absdiff{17}{8.9} \\
        & Claude Code & \absdiff{11}{0.20} & \absdiff{23}{15.6} & \absdiff{11}{3.3} & \absdiff{24}{1.00} & \absdiff{21}{12.7} & \absdiff{31}{23.3} & \absdiff{11}{0.20} & \absdiff{32}{24.6} & \zerodiff{0.0} & \absdiff{11}{0.20} & \absdiff{17}{9.3} & \absdiff{17}{8.9} \\
        & \ourtool{}  & \absdiff{9}{0.04} & \absdiff{10}{2.0} & \zerodiff{0.0} & \zerodiff{0.0} & \zerodiff{0.0} & \zerodiff{0.0} & \absdiff{22}{0.87} & \absdiff{21}{12.7} & \absdiff{18}{10.0} & \absdiff{13}{0.30} & \absdiff{11}{3.5} & \absdiff{11}{3.3} \\
        \midrule

        \multirow{3}{*}{Claude Sonnet 4.6}
        & Cursor      & \absdiff{13}{0.33} & \absdiff{9}{0.7} & \absdiff{11}{3.3} & \absdiff{14}{0.40} & \absdiff{16}{8.0} & \absdiff{11}{3.3} & \absdiff{13}{0.33} & \absdiff{19}{10.7} & \absdiff{15}{6.7} & \absdiff{9}{0.09} & \absdiff{9}{1.1} & \zerodiff{0.0} \\
        & Claude Code & \absdiff{10}{0.10} & \absdiff{9}{0.6} & \zerodiff{0.0} & \absdiff{13}{0.30} & \absdiff{11}{3.4} & \zerodiff{0.0} & \absdiff{11}{0.17} & \absdiff{9}{0.6} & \zerodiff{0.0} & \absdiff{10}{0.12} & \absdiff{9}{1.2} & \zerodiff{0.0} \\
        & \ourtool{}  & \absdiff{11}{0.20} & \absdiff{9}{1.3} & \absdiff{11}{3.4} & \absdiff{10}{0.13} & \absdiff{12}{4.0} & \zerodiff{0.0} & \absdiff{12}{0.27} & \absdiff{15}{6.6} & \absdiff{11}{3.3} & \absdiff{11}{0.20} & \absdiff{8}{0.4} & \zerodiff{0.0} \\
        \midrule

        \multirow{3}{*}{Claude Haiku 4.5}
        & Cursor      & \absdiff{10}{0.10} & \zerodiff{0.0} & \zerodiff{0.0} & \absdiff{21}{0.80} & \absdiff{29}{21.4} & \absdiff{11}{3.3} & \absdiff{12}{0.23} & \absdiff{9}{0.6} & \zerodiff{0.0} & \absdiff{13}{0.31} & \absdiff{10}{2.2} & \absdiff{9}{1.1} \\
        & Claude Code & \absdiff{22}{0.87} & \absdiff{9}{0.7} & \zerodiff{0.0} & \absdiff{35}{1.67} & \absdiff{22}{14.6} & \absdiff{18}{10.0} & \absdiff{33}{1.53} & \absdiff{21}{12.7} & \absdiff{18}{10.0} & \absdiff{30}{1.36} & \absdiff{17}{9.3} & \absdiff{15}{6.6} \\
        & \ourtool{}  & \absdiff{10}{0.10} & \absdiff{9}{0.7} & \zerodiff{0.0} & \absdiff{10}{0.10} & \zerodiff{0.0} & \zerodiff{0.0} & \absdiff{16}{0.50} & \absdiff{9}{0.7} & \absdiff{11}{3.3} & \absdiff{12}{0.24} & \zerodiff{0.0} & \absdiff{9}{1.1} \\
        \midrule

        \multirow{2}{*}{GPT 5.4}
        & Cursor      & \absdiff{9}{0.04} & \absdiff{19}{11.4} & \absdiff{18}{10.0} & \absdiff{16}{0.50} & \absdiff{20}{12.0} & \absdiff{18}{10.0} & \absdiff{11}{0.17} & \absdiff{15}{6.6} & \absdiff{11}{3.4} & \absdiff{10}{0.10} & \absdiff{14}{5.6} & \absdiff{13}{5.5} \\
        & \ourtool{}  & \absdiff{17}{0.56} & \absdiff{10}{2.0} & \absdiff{11}{3.3} & \absdiff{20}{0.77} & \absdiff{27}{19.4} & \absdiff{21}{13.3} & \absdiff{8}{0.03} & \absdiff{12}{4.0} & \zerodiff{0.0} & \absdiff{9}{0.08} & \absdiff{15}{7.3} & \absdiff{11}{3.4} \\
        \midrule

        \multirow{2}{*}{GPT 5.4 mini}
        & Cursor      & \absdiff{14}{0.40} & \absdiff{17}{8.7} & \absdiff{11}{3.4} & \absdiff{20}{0.74} & \absdiff{34}{26.0} & \absdiff{38}{30.0} & \absdiff{12}{0.23} & \absdiff{11}{3.4} & \absdiff{15}{6.7} & \absdiff{15}{0.45} & \absdiff{22}{13.6} & \absdiff{21}{13.4} \\
        & \ourtool{}  & \absdiff{9}{0.07} & \absdiff{11}{2.7} & \absdiff{15}{6.6} & \absdiff{19}{0.67} & \absdiff{28}{20.0} & \absdiff{25}{16.7} & \absdiff{9}{0.06} & \absdiff{16}{8.0} & \absdiff{15}{6.7} & \absdiff{11}{0.18} & \absdiff{16}{8.0} & \absdiff{9}{1.2} \\
        \midrule

        \multirow{1}{*}{Qwen 3.5 9B}
        & \ourtool{}  & \zerodiff{0.0} & \absdiff{13}{4.7} & \absdiff{15}{6.7} & \absdiff{14}{0.37} & \absdiff{23}{15.3} & \absdiff{18}{10.0} & \zerodiff{0.0} & \zerodiff{0.0} & \zerodiff{0.0} & \absdiff{10}{0.12} & \absdiff{14}{6.2} & \absdiff{14}{5.6} \\

        \bottomrule
    \end{tabular}%
    \caption{Absolute divergence between LLM and human expert scoring results, computed as $|\mathrm{LLM} - \mathrm{human expert}|$. Pass@1 and pass@5 divergences are in percentage points (pp). Darker purple indicates larger divergence.}
    \label{tab:comparison_accuracy_absdiff}
\end{table*}

\subsubsection{Limitations of Council of LLMs as a Judge}
When assessing the scoring discrepancies between the evaluation strategies, we find the following notable differences. In statistical questions, we find that the LLM-Council results in higher scores relative to human expert scoring, particularly within the Cursor and Claude-Code agents, with the average score increasing by (+0.3-1.67) across all underlying LLMs. In contrast, the results from \ourtool{} remained largely consistent with the human expert evaluations, with 3 of the 4 LLM configurations yielding equivalent or degraded scores. We find that this is largely due to the way the units' correctness is assessed relative to the numerical value itself. For instance, LLM-Council was more inclined to award partial credit for a numerical value, even if the unit was either wrong or missing (e.g., Watts, meters, seconds), since the scalar value itself would still be consistent. In contrast, for unit inaccuracies, human experts were more likely to score harshly, given the incorrect nature of the response, regardless of the numerical similarity.

Given these limitations in the LLM-Council scoring approach, we determine that the human expert results are most representative of the agent's capabilities within the EDA benchmark tasks and are used in the paper's primary findings.

\end{document}